\pdfoutput=0
\documentclass[mathpazo]{cicp}

\usepackage{epstopdf} 
\usepackage{caption}
\usepackage{url}
\usepackage{wrapfig}
\usepackage[T1]{fontenc}
\usepackage[utf8]{inputenc}
\usepackage{color}
\usepackage{amsmath, amssymb, mathtools, latexsym, natbib}
\usepackage{psfrag,epsfig,amsfonts,latexsym,amsthm,,amscd,url }
\usepackage{bm}
\usepackage{appendix}
\usepackage{subcaption}
\usepackage{booktabs} 
\usepackage{algorithm}
\usepackage{algpseudocode}
\usepackage{ragged2e}
\usepackage[colorlinks=true,linkcolor=blue,citecolor=blue, linktocpage=true]{hyperref}
\allowdisplaybreaks[4]

\usepackage{overpic}

\renewcommand{\phi}{\varphi}

\newcommand{\grad}{\nabla}

\newcommand{\dd}{{\rm d}}
\newcommand{\x}{{\bm x}}
\newcommand{\X}{{\bm X}}

\theoremstyle{plain}

\theoremstyle{definition}

\theoremstyle{remark}

\newcommand{\ub}{\mathbf{u}}

\newcommand{\BE}{\begin{equation}}
\newcommand{\EE}{\end{equation}}
\newcommand{\BEN}{\begin{equation*}}
\newcommand{\EEN}{\end{equation*}}
\newcommand{\BAL}{\begin{align}}
\newcommand{\EAL}{\end{align}}
\newcommand{\BAN}{\begin{align*}}

\DeclareMathOperator*{\argmin}{arg\,min}

\begin{document}
\title{Learning Generalized Diffusions using an Energetic Variational Approach}


\author[Lu Y et.~al.]{Yubin Lu\affil{1}\comma\corrauth,
      Xiaofan Li\affil{1}, Chun Liu\affil{1}, Qi Tang\affil{2}, and Yiwei Wang\affil{3}}
\address{\affilnum{1}\ Department of Applied Mathematics, Illinois Institute of Technology, Chicago, IL 60616, United States \\
          \affilnum{2}\ School of Computational Science and Engineering, Georgia Institute of Technology, Atlanta, GA 30332, United States \\
          \affil{3}\ Department of Mathematics, University of California, Riverside, Riverside, CA 92521, United States}
\emails{{\tt ylu117@illinoistech.edu} (Y.~Lu), {\tt lix@illinoistech.edu} (X.~Li),  {\tt cliu124@illinoistech.edu} (C.~Liu),
         {\tt qtang@gatech.edu} (Q.~Tang), {\tt yiweiw@ucr.edu} (Y.~Wang)}

\begin{abstract}
Extracting governing physical laws from computational or experimental data is crucial across various fields such as fluid dynamics and plasma physics.
Many of those physical laws are dissipative due to fluid viscosity or plasma collisions.
For such a dissipative physical system, we propose a framework to learn the corresponding laws of the systems based on their energy-dissipation laws,
assuming either continuous data (probability density) or discrete data (particles) are available. Our methods offer several key advantages, including their robustness to corrupted/noisy observations, their easy extension to more complex physical systems, and the potential to address higher-dimensional systems. We validate our approaches through representative numerical examples and carefully investigate the impacts of data quantity and data property on model discovery.
\end{abstract}


\maketitle

\section{Introduction}

Constructing surrogate models that approximate the behavior of physical systems and discovering physical laws, often represented by nonlinear partial differential equations (PDEs), are two major data-driven approaches that can help us better understand complex natural phenomena. As a concrete example, generalized diffusion, a type of mechanical process involving a conserved quantity, can be described by an energy-dissipation law. Generalized diffusion encompasses a wide range of models across various fields, including the Fokker-–Planck equation(FPE) \cite{risken1996fokker}, the porous medium equation \cite{PME}, and the Poisson–Nernst–Planck equation \cite{PNP,LiPNP,LiuPNP}. Constructing surrogate models and discovering physical laws for such systems can serve as effective alternatives or complements to costly equation-based state-of-the-art methods, enabling tasks such as prediction, optimal control, and uncertainty quantification.

One of the most widely used methods for discovering physical laws is the physics-informed neural network (PINN) \cite{PINN}. The idea of PINN is to train neural networks using a loss function based on the underlying partial differential equation and noisy observation data. 
This approach can be traced back to at least the works in \cite{Yannis1,Yannis2,Yannis3}. 
Another powerful approach for extracting governing physical laws from data is a sparse identification of nonlinear dynamical systems (SINDy) \cite{SINDy}.
SINDy has gained popularity due to its interpretability and computational efficiency. The SINDy framework is motivated by the pioneer work \cite{bongard2007automated, schmidt2009distilling}, which uses symbolic regression to recover physical equations from data. 
Later, a weak-form version of SINDy was developed for learning PDEs \cite{weakSINDy, weakSINDy4PDE} and extended to cover mean-field equations \cite{weakSINDy4meanfield} and Hamiltonian systems \cite{weakSINDy4Hamiltonian}.
Koopman operator theory is also used to establish various data-driven analysis for complex dynamics \cite{Koopman1,EDMD,Koopman2}. Nonparametric regression techniques for learning interaction kernels \cite{KernelLearning1,KernelLearning2,KernelLearning3,KernelLearning4,KernelLearning5,KernelLearning6,Ding2022} are developed for various equations. Flow maps \cite{flowMap1,flowMap2,flowMap3} and kernel flows \cite{kernelFlow1} for learning dynamical systems are introduced. Probabilistic/statistical methods, including Bayesian inferences, maximum likelihood methods, Gaussian processes, kernel methods, and Wasserstein distances, are introduced to learn stochastic dynamical systems \cite{Felix,VI4SDE,LiuSWasserstein, GP4PDEs1,GP4PDEs2,Kernel4PDEs}.  More recently, in order to maintain the physical properties (e.g., invariant quantities for conserved systems or dissipation rates for dissipative systems) of the original system while learning the system, various structure-preserving learning strategies are developed to learn Hamiltonian systems \cite{SympNets,HenonMaps, MTao,Hamiltonian1,Hamiltonian2,Hamiltonian3,Hamiltonian4,Hamiltonian5,SymplecticRNN,MLforIrreversibleProcess1}, energy dissipative systems \cite{dissipativeNN2,OnsagerNet,stat-PINNs}, {and, more generally, metriplectic systems \cite{metriplectic1,MLforIrreversibleProcess2}.}

However, most existing work establishes the learning framework based on the corresponding governing equations, such as (stochastic) ordinary differential equations (ODEs/SDEs) and partial differential equations (PDEs). The resulting models, however, may fail to preserve fundamental physical principles, such as consistency with thermodynamic laws. Recently, there are growing interesting of learning thermodynamically consistent physical model from variational principles, such as  General Equation for Non-Equilibrium Reversible-Irreversible Coupling (GENERIC) formalism \cite{hernandez2021structure, zhang2022gfinns} and Onsager principle \cite{huang2022variational, OnsagerNet}.  These variational principles model complex physical processes by accounting for both energy conservation (in reversible processes) and energy dissipation (in irreversible processes). The key idea behind these variational principle-based learning approaches is to parameterize the physical quantities in the energy-dissipation law using neural networks while constructing loss functions based on equations derived from these principles.

The variational formulation of loss functions has gained increasing attention recently due to its robustness to corrupted or noisy observations. For example, several works have focused on leveraging the weak form of PDEs to construct loss functions for learning the solution of PDEs in forward problems or identifying coefficients in inverse problems \cite{DGM,neuralGalerkin,weakPINNs,variationalPINN,weakSINDy, weakSINDy4PDE,weakSINDy4meanfield, Phase-field_DeepONet}. However, most existing work relies on the careful selection of test functions, which can be challenging or even infeasible for high-dimensional problems. Additionally, we note a concurrent independent effort that also aims to address high-dimensional problems by introducing self-test loss functions for learning weak-form operators and gradient flows \cite{Lu_self-test}. Moreover, the authors in \cite{EntropyInformed} proposed an entropy-informed learning framework.

The goal of this work is to propose a new learning framework based on the energy-dissipation laws of the target physical systems directly, without relying on the governing equations. Our proposed methods offer several benefits, including robustness to corrupted or noisy observations, straightforward extensions to more general physical systems, and the potential to handle higher-dimensional systems. 
Moreover, our approach can learn the full dynamics of the system using observations at only three time instances, which not only reduces data requirements but also enables efficient modeling in scenarios where long-time trajectory data are difficult to obtain. 
While loss functions formulated in the weak form of governing equations are generally more robust to noise than those based on strong formulations, they may struggle to uniquely capture local information and can be challenging to construct for complex systems. 
Our approach constructs the loss function directly from the energy-dissipation law, enabling effective learning of PDE solutions in forward problems and accurate identification of model parameters in inverse problems, without relying on the explicit form of the governing equations.

In this work, we focus on learning the potential function and noise intensity in one- or two-dimensional generalized diffusions to illustrate our method and explore its performance under different settings. While extending the approach to higher-dimensional problems and other physical systems is straightforward, we leave this direction for future work.
The rest of the paper is organized as follows. Section \ref{sec:Formulation} provides a brief introduction to the energetic variational approach for generalized diffusions. In Section \ref{sec:learning}, we propose a framework for learning the governing laws of the systems based on their energy-dissipation laws, using either continuous data (probability density) or discrete data (SDE particles). Section \ref{sec:examples} presents several representative examples to validate the performance of our methods. Finally, we conclude with a brief discussion in Section \ref{sec:discussion}.

\section{Formulation}\label{sec:Formulation}

Before proposing the learning framework, we briefly introduce the energetic variational approach (EnVarA for short) \cite{EnVarA1} for generalized diffusions, which plays an important role in our proposed learning frameworks in the next section. 

Motivated by non-equilibrium thermodynamics, particularly the seminal work of Rayleigh \cite{Rayleigh} and Onsager \cite{Onsager1,Onsager2}, an isothermal and mechanically-closed complex system can be described by an energy-dissipation law
\begin{equation}\label{eqn:original energylaw}
    \frac{d}{dt}E^{\rm total}=-\Delta\leq 0,
\end{equation}
where $E^{{\rm total}}$ is the sum of the kinetic energy $\mathcal{K}$ and the Helmholtz free energy $\mathcal{F}$, and $\Delta$ is the rate of energy dissipation.  Based on the energy-dissipation law \eqref{eqn:original energylaw}, EnVarA is a unique, well-defined way to derive the dynamics of the underlying system using the least action principle (LAP) and the maximum dissipation principle (MDP). To be more specific, for the Hamiltonian part of the system, one can employ the LAP, taking variation of the action functional $\mathcal{A}({\bm x}) = \int_0^T \left( \mathcal{K} - \mathcal{F} \right) \dd t$ with respect to ${\bm x}$ (the trajectory in Lagrangian coordinates) \cite{EnVarA1, arnol2013mathematical},  to derive the conservative force, i.e.,
$\delta \mathcal{A} =  \int_{0}^T \int_{\Omega} ({\rm force}_{\text{iner}} - {\rm force}_{\text{conv}})\cdot \delta {\bm x}  ~ \dd \x \dd t.$ Here, $\Omega$ could be a bounded or unbounded domain of $\x$, and ${\rm force}_{\text{iner}}$ and ${\rm force}_{\text{conv}}$ are inertial force and conservative force respectively.  For the dissipation part, one can apply the MDP, taking the variation of the Onsager dissipation functional $\mathcal{D}$ with respect to the ``rate'' $\dot{\x}$ ($\dot{\x}$ is the derivative of the trajectory $\x$ with respect to time $t$), to derive the dissipative force, i.e., $\delta \mathcal{D}  = \int_{\Omega} {\rm force}_{\text{diss}} \cdot \delta \dot{\x}~ \dd \x$, where the dissipation functional $\mathcal{D} = \frac{1}{2} \triangle$ in the linear response regime \cite{Onsager2} and $ {\rm force}_{\text{diss}}$ is the dissipative force. 
Subsequently, the force balance condition connects the conservative force and the dissipation force providing the evolution equation of the studied system   
\begin{align}
    \frac{\delta\mathcal{D}}{\delta \dot{\x} }=\frac{\delta\mathcal{A}}{\delta\bm x}.
\end{align}
The EnVarA has been successfully applied to build various mathematical models in physics, chemical engineering, and biology \cite{EnVarA_review}.

\paragraph{\bf Generalized Diffusion}
Let us consider the following random process
    \begin{equation}\label{eqn:generalized_diff}
        \dd \bm X_t=\bm a(\bm X_t) \dd t +\sigma(\bm X_t)  \dd W_t,
    \end{equation}
    where $W_t$ is a standard $n$-dimensional Brownian motion,  $\bm X_t$ and $\bm a$ are two $n$-dimensional vectors denoting the state variable at time $t\in\mathbb{R}^+\cup\{0\}$ and the drift coefficient respectively, and the noise intensity $\sigma$ is a \textbf{scalar} function. If the stochastic integral of \eqref{eqn:generalized_diff} is interpreted as {\bf backward It\^{o} integral} \cite{russo1993forward}, one may obtain the following Fokker--Planck equation (See (c) in Remark \ref{rmk:energyLaw}):
    \begin{equation}\label{eqn:FPE_generalizedDiff}
        f_t+\nabla\cdot(\bm af)=\frac{1}{2}\nabla\cdot\left(\sigma^2\nabla f\right),
    \end{equation}
where $f({\bm x}, t)$ is the probability density function of the state variable $\bm X_t$.

According to the fluctuation-dissipation theorem \cite{kubo1966fluctuation}, the convection coefficient is constrained by
    \begin{equation}\label{eqn:FD theorem}
        \bm a =-\frac{1}{2}\sigma^2\nabla\psi,
    \end{equation}
where $\psi$ is the potential function and $\sigma$ is the noise intensity. 

The fluctuation-dissipation theorem ensures the existence of an energy-dissipation law associated with the Fokker–Planck equation (\ref{eqn:FPE_generalizedDiff}). It can be shown that the Fokker--Planck equation (\ref{eqn:FPE_generalizedDiff}) with the condition (\ref{eqn:FD theorem}) satisfies energy-dissipation law:
\begin{equation}\label{eqn:EnergyLaw_generalizedDiff}
        \frac{d \mathcal{F}[f]}{dt}=-\int_\Omega\frac{f}{\sigma^2/2}|\bm u|^2d\bm x,
    \end{equation}
    along with the continuity equation of the probability density
    \begin{equation} \label{eqn:diffusion}
    f_t + \nabla \cdot (f {\bm u}) = 0.
    \end{equation}
    Here, ${\bm u}$ is a certain average velocity of all stochastic trajectories and $\mathcal{F}[f]$ is the free energy given by
    \begin{equation}\label{eqn:energy}
      {\bm u} = -\frac{\sigma^2}{2} \nabla (\ln f + \psi ), \quad  \mathcal{F}[f] := \int_\Omega \left[f\ln f+\psi f\right]d\bm x\ .
    \end{equation}

From a modeling perspective, one can derive the evolution equation \eqref{eqn:FPE_generalizedDiff} from the energy-dissipation law (\ref{eqn:EnergyLaw_generalizedDiff}) by the general framework of EnVarA \cite{EnVarA1}. Note that
\begin{equation}
      \mathcal{K}=0, \qquad \mathcal{F} = \int_\Omega \left[f\ln f+\psi f\right]d\bm x,\qquad 
      \mathcal{D}=\frac{1}{2}\int_\Omega\frac{f}{\sigma^2/2}|\bm u|^2d\bm x.
\end{equation} 
To apply the LAP, we need first introduce the concept of flow map $\x(\X, t)$, defined through
\begin{equation} \label{eqn:ODE_flow_map}
 \begin{cases}
 & \frac{\dd}{\dd t} \x(\X, t) = {\bm u}(\x(\X, t), t), \\
 & \x(\X, 0) = \X,\\
 \end{cases}
\end{equation}
for a given velocity field ${\bm u}$.
Here $\X$ is the Lagrangian coordinate and $\x$ is the Eulerian coordinates. For fixed $\X$, $\x(\X, t)$ can be interpreted as the trajectory of the particle that is initially located at $\X$. Due to the mass conservation, $f(\x, t)$ can be viewed as the function of the flow map $x(\X, t)$, as
\begin{equation}
  f(\x, t) =  f_0 (\X) / \det ( \nabla_{X} \x(\X, t) )
\end{equation}
where $f_0(\X)$ is the initial density. Consequently, one can take the variational of the action functional with respect to the flow map $\x(\X, t)$.
The final force balance equation is given by 
\begin{equation}
{\bm u}(\x, t) = -\left(\frac{\sigma^2}{2}\nabla\ln f+\frac{\sigma^2}{2}\nabla\psi\right),
\end{equation}
which is the velocity derived from the energy-dissipation law (\ref{eqn:EnergyLaw_generalizedDiff}).
Combining with the continuity equation~\eqref{eqn:diffusion}, one can obtain the 
Fokker--Planck equation \eqref{eqn:FPE_generalizedDiff} with ${\bm a}$ given by (\ref{eqn:FD theorem}). 
An advantage of deriving the governing equation from an energy-dissipation law is that the resulting system is automatically thermodynamically consistent, meaning it satisfies the fluctuation-dissipation theorem in this case.
We refer the interested reader to \cite{EnVarA1, hu2024energetic} and the references therein for more details.

\begin{remark}\label{rmk:energyLaw}
    Different interpretations of the stochastic integral of \eqref{eqn:generalized_diff} leads to different energy-dissipation law \cite{EnVarA1}. To be more specific, writing a Taylor expansion of probability distribution function $f(\bm x,t)$, one may obtain the following PDEs \cite{EnVarA1}:
    
    (a) $f_t+\nabla\cdot(\bm af)=\frac{1}{2}\Delta(\sigma^2f)$ if using It\^{o} integral,

    (b) $f_t+\nabla\cdot(\bm af)=\frac{1}{2}\nabla\cdot[\sigma\nabla(\sigma f)]$ if using Stratonovich integral,

    (c) $f_t+\nabla\cdot(\bm af)=\frac{1}{2}\nabla\cdot[\sigma^2\nabla f]$ if using backward It\^{o} integral, yielding PDE with self-adjoint diffusion term. If the convection coefficient satisfies the fluctuation-dissipation theorem~\eqref{eqn:FD theorem}, i.e. $\bm a =-\frac{1}{2}\sigma^2\nabla\psi$,
then the above PDEs may be obtained from variation of the following energy laws respectively

    (a) $\frac{d}{dt}\int\left[f\ln(\frac{1}{2}\sigma^2f)+\psi f\right]d\bm x=-\int\frac{f}{\sigma^2/2}|\bm u|^2d\bm{x}$,

    (b) $\frac{d}{dt}\int\left[f\ln(\sigma f)+\psi f\right]d\bm x=-\int\frac{f}{\sigma^2/2}|\bm u|^2d\bm{x}$,

    (c) $\frac{d}{dt}\int\left[f\ln f+\psi f\right]d\bm x=-\int\frac{f}{\sigma^2/2}|\bm u|^2d\bm{x}$
along with the mass conservation~\eqref{eqn:diffusion}.
\end{remark}

\begin{remark}
    In the current study, we establish the learning framework based on the expression (c) in Remark \ref{rmk:energyLaw}. Therefore, the SDE \eqref{eqn:generalized_diff} is interpreted as a backward It\^{o} integral correspondingly. The reason for choosing (c) is that both sides of the first two expressions, (a) and (b), depend on the noise intensity $\sigma$, which exacerbates the ill-posedness of the problem, as we must balance both sides during the training process. To simulate the backward It\^{o} SDE (\ref{eqn:generalized_diff}), we rewrite it as a standard It\^{o} SDE \begin{equation}\label{Ito_SDE_1}
    d\bm X_t=\left[\bm a(\bm X_t)+\nabla\left(\frac{1}{2}\sigma^2(\bm X_t)\right)\right]dt+\sigma(\bm X_t)dW_t
    \end{equation}
    in \eqref{eqn:generalized_diff} and apply the Euler–Maruyama scheme \cite{weinan2021applied}. It should be noted that there is a slight abuse of notation here. The stochastic integral $\sigma(\bm X_t)dW_t$ in (\ref{Ito_SDE_1}) is interpreted as an It\^{o} integral, whereas the stochastic integral $\sigma(\bm X_t)dW_t$ in \eqref{eqn:generalized_diff} is interpreted as a backward It\^{o} integral.     
\end{remark}

\section{Learning framework}\label{sec:learning}
In this section, we propose a learning framework designed to identify (partial) dynamics of the generalized diffusion equation \eqref{eqn:generalized_diff}, using two types of data: continuous data (e.g., probability densities) and discrete data (e.g., particle trajectories).

We assume that the generalized diffusion satisfies the fluctuation-dissipation theorem, which relates the noise intensity to the drift term by $\bm a =-\frac{1}{2}\sigma^2\nabla\psi$ in \eqref{eqn:generalized_diff}. Our goal is to identify the potential function $\psi$ and/or the noise intensity $\sigma^2$ (in what follows, we refer to both $\sigma$ and $\sigma^2$ as noise intensity) of the generalized diffusion \eqref{eqn:generalized_diff} from data. Furthermore, we investigate how the nature of the available data influences the learning task, and accordingly develop different learning strategies suited to each data type. 

The proposed framework is based on the energy functional \eqref{eqn:energy}. Thanks to the fluctuation-dissipation theorem, the system \eqref{eqn:generalized_diff} or \eqref{eqn:FPE_generalizedDiff} satisfies the energy-dissipation law \eqref{eqn:EnergyLaw_generalizedDiff}. Therefore, we can learn the potential function $\psi$ and/or the noise intensity $\sigma^2$ by checking against the energy-dissipation law \eqref{eqn:EnergyLaw_generalizedDiff}.

Our loss function is constructed directly from the energy-dissipation law \eqref{eqn:EnergyLaw_generalizedDiff}, rather than from the governing equations. This approach offers several advantages. First, it relies solely on an energy-dissipation law, bypassing the need for information from the governing equations. Second, since the energy-dissipation law is expressed in an integral (weak) form, it imposes weaker regularity requirements on the density function, which is likely to be more robust to corrupted/noisy observations compared to loss functions based on governing equations. Third, the integral form of the loss function has the potential to be extended to handle higher dimensional problems efficiently, such as through the use of particle methods. 

In this section the energy-dissipation law is expressed in terms of the probability density function $f$, as the most straightforward way. For simplicity, we illustrate our methods by assuming the noise intensity $\sigma^2$ is known and focus on learning the potential function $\psi$. Alternatively, we could also learn  the noise intensity $\sigma^2$ while assuming the potential function $\psi$ is known. Here we  let the unknown potential function $\psi(\bm x)$ be approximated by a neural network ${\psi}_{nn}(\bm x;\theta)$.

\subsection{Density-based Method}\label{sec:density-based}

Since the free energy $E$ and the velocity $\bm u$ in \eqref{eqn:energy} and the dissipation rate in \eqref{eqn:EnergyLaw_generalizedDiff} are expressed in terms of the probability density function $f$, it is most straightforward to compute the loss function based on the density data $f$. 
The observation dataset, consisting of probability density values at \textbf{three consecutive time instances} with a fixed time interval \(\delta t\), is denoted by
\[
\left\{ \left( f_j(\bm{x}_{i}, t_1), f_j(\bm{x}_{i}, t), f_j(\bm{x}_{i}, t_2) \right) \right\}_{i,j=1}^{N,M},
\]
where \(t_1 = t - \delta t\) and \(t_2 = t + \delta t\). Here, \(\{\bm{x}_{i}\} \subset \Omega\) are the \(N\) uniform grid points with spatial resolution \(\Delta x\) for each \(j\), and \(M\) is the number of instances generated from \(M\) different initial distributions.

The free energy \eqref{eqn:energy} at time $t$ can be approximated by the following Riemann sum approximation
\begin{align}\label{eqn:discrete_energy_Riemann}
    E_j^N(t,\theta) = \sum\limits_{i=1}^N \left[f_j(\bm x_{i},t)\ln f_j(\bm x_{i},t) + \psi_{nn}(\bm x_{i};\theta)f_j(\bm x_{i},t)\right]\Delta x.
\end{align}

Since the density function data is assumed to be available in this case, we construct the loss function based on the original energy-dissipation law \eqref{eqn:EnergyLaw_generalizedDiff}
\begin{equation}\label{eqn:loss_density}
    \theta^* = \argmin\limits_{\theta} \sum\limits_{j=1}^{M}\lambda(j)\left\|\frac{E_j^N(t_{2};\theta)-E_j^N(t_{1};\theta)}{t_{2}-t_1}+\Delta x\sum\limits_{i=1}^N\frac{f_j(\bm x_{i},t)}{\sigma^2/2}\left|\frac{\sigma^2}{2}\nabla\ln f_j(\bm x_{i},t)+\frac{\sigma^2}{2}\nabla\psi_{nn}(\bm x_{i};\theta)\right|^2\right\|^2,
\end{equation}

\noindent where $t_1=t-\delta t$ and $t_2=t+\delta t$ for a given observation time step size $\delta t$ and $\lambda$ is an user-defined weighting function. We note that, if the training data for $f$ were obtained by solving the Fokker--Planck equation \eqref{eqn:FPE_generalizedDiff}, it would be computationally expensive in high dimensions.
\begin{remark}
    The loss function~\eqref{eqn:loss_density} is in an integral/summation form, which has lower regularity requirements compared to the corresponding PDE~\eqref{eqn:FPE_generalizedDiff}. This integral form is expected to enhance the robustness of the proposed density-based method, particularly when the density function is not smooth enough or the observed density function is affected by polluted observations. We will present a simple comparison between our EnVarA-based method and a simplified PDE-based method in the numerical examples in the next section. However, this does not imply that our EnVarA-based method outperforms PDE-based methods in all scenarios, as PDE-based methods can offer more detailed local information. Therefore, our goal is not to compete with state-of-the-art methods, but rather to present an alternative approach that may be advantageous in certain situations.
\end{remark}
\begin{remark}
    We can learn the potential function $\psi$ by minimizing the loss function~\eqref{eqn:loss_density} given the noise intensity $\sigma^2$. Conversely, we can also learn the noise intensity $\sigma^2$ if the potential function $\psi$ is provided. Indeed, these two learning tasks have different data requirements for the training data $\{(f_j(\bm x_{i},t_1),\allowbreak f_j(\bm x_{i},t),\allowbreak f_j(\bm x_{i},t_2))\}_{i,j=1}^{N,M}$ in the proposed density-based method. By noticing that~\eqref{eqn:loss_density} is a weak-form loss function, the two learning problems are ill-posed in general. When learning the potential function $\psi$, if the training data are stationary, the approximation of $dE/dt$ in the loss function~\eqref{eqn:loss_density} becomes zero. As a result, the originally ill-posed problem transforms into a well-posed one, meaning that~\eqref{eqn:loss_density} serves as a point-wise loss function in this case. In contrast, stationary training data are not suitable for learning the noise intensity $\sigma^2$, since the approximation of $dE/dt$ remains zero, making zero a minimizer of the loss function. We will further explore this in the next section through numerical examples.

\end{remark}
\begin{remark}\label{remark:2nd-order}
    {The free energy of the system \eqref{eqn:generalized_diff} decays exponentially over time, particularly in the initial stage, the derivative ($dE/dt$) is large. We found first-order schemes lack sufficient accuracy, which potentially impacts the performance of our method. Therefore, we use a second-order scheme here instead of the forward Euler scheme to achieve more accurate derivative ($dE/dt$) estimates. To do so, we collect training data $\{(f_j(\bm x_{i},t_1),\allowbreak f_j(\bm x_{i},t), \allowbreak f_j(\bm x_{i},t_2))\}_{i,j=1}^{N,M}$ at three time instances to compute the derivative $dE/dt$ using a more accurate finite difference scheme, specifically a second-order central difference approximation.}
\end{remark}

\subsection{Particle-to-density method}\label{sec:particle-to-density}
Next, we consider the case where the probability function $f$ corresponding to the state variable is not readily available. 
Solving a high-dimensional Fokker--Planck equation using a continuous representation of $f$ faces the curse of dimensionality, which becomes less practical.
Therefore, we propose an alternative way to establish the learning framework here. 

Suppose that we can access particle data that satisfy the SDE \eqref{eqn:generalized_diff} instead of the probability density function $f$. The observation dataset, consisting of particle trajectories at \textbf{three consecutive time instances} with a fixed time interval \(\delta t\), is denoted by
\[
\left\{ \left( \bm{x}_{i,j}(t_1), \bm{x}_{i,j}(t), \bm{x}_{i,j}(t_2) \right) \right\}_{i,j=1}^{N_s,M},
\]
where \(t_1 = t - \delta t\) and \(t_2 = t + \delta t\).
Here, \(N_s\) denotes the sample size representing the distribution function. The parameter \(M\) corresponds to the number of instances generated from \(M\) different initial conditions.

One can approximate the probability density function $f$ using particle data $\{(\bm x_{i,j}(t_1),\allowbreak\bm x_{i,j}(t),\bm x_{i,j}(t_2))\}_{i,j=1}^{N_s,M}$, denoted by $f^{N_s}$, so that the loss function \eqref{eqn:loss_density} can be computed as in the density-based method. For each $j$, the underlying density $f_j^{N_s}(x, t)$ can be estimated from the particle samples $\{ {\bm x}_{i, j} \}_{i=1}^{N_s}$ using various methods.
In this work, we use the kernel density estimation (KDE) method \cite{KDE1,KDE2} to approximate the density function. It is worth noting that selecting the bandwidth in KDE is a delicate task, particularly for high-dimensional density functions. As an alternative, one can use normalizing flows \cite{NFs1,NFs2,Lu2} to estimate the density from particle data, as this estimation is carried out during a pre-training step.

Subsequently, the loss function for the particle-to-density method can be obtained by replacing $f$ with $f^{N_s}$ in the loss function \eqref{eqn:loss_density} of the density-based method 
\begin{equation}\label{eqn:loss_particle2density}
    \theta^* = \argmin\limits_{\theta} \sum\limits_{j=1}^{M}\lambda(j)\left\|\frac{E_j^{N_s}(t_{2};\theta)-E_j^{N_s}(t_{1};\theta)}{t_{2}-t_1}+\Delta x\sum\limits_{i=1}^{N_s}\frac{f_j^N(\bm x_{i},t)}{\sigma^2/2}\left|\frac{\sigma^2}{2}\nabla\ln f_j^{N_s}(\bm x_{i},t)+\frac{\sigma^2}{2}\nabla\psi_{nn}(\bm x_{i};\theta)\right|^2\right\|^2,
\end{equation}
\noindent where the energy is
\begin{align}\label{eqn:discrete_energy_Riemann_particle2density}
    E_j^{N_s}(t,\theta) = \sum\limits_{i=1}^{N_s} \left[f_j^{N_s}(\bm x_{i},t)\ln f_j^{N_s}(\bm x_{i},t) + \psi_{nn}(\bm x_{i};\theta)f_j^{N_s}(\bm x_{i},t)\right]\Delta x.
\end{align}

Compared with the density-based method, the particle-to-density method gives a less accurate learning framework since we need to approximate the density function using particle data. However, as a reward at the cost of losing accuracy, we can obtain training datasets efficiently, especially in high dimensions, since we can solve the SDE \eqref{eqn:generalized_diff} instead of solving the Fokker--Planck equation \eqref{eqn:FPE_generalizedDiff}.

To clearly present the proposed EnVarA-based learning methods, we provide a brief algorithm below. See Algorithm \ref{Alg:EnVarA}.

\begin{algorithm}
\caption{Learning generalized diffusion using EnVarA}
\label{Alg:EnVarA}
\begin{minipage}{\linewidth}
\raggedright
\begin{itemize}
    \item Particle data of three time steps $\{(\bm x_{i,j}(t_1),\bm x_{i,j}(t),\bm x_{i,j}(t_2))\}_{i,j=1}^{N_s,M}$ or probability density functions of three time steps $\{(f_j(\bm x_{i},t_1), f_j(\bm x_{i},t), f_j(\bm x_{i},t_2))\}_{i,j=1}^{N,M}$ are given for training. For the former case, one can approximate the probability density function from particle data using KDE, denoted by $f^{N_s}$. 
    \item Optimize the loss function \eqref{eqn:loss_density} or \eqref{eqn:loss_particle2density} to find the `‘best’' parameters of the neural networks.
    \item Reconstruct the learned potential function $\psi_{nn}$ or the noise intensity $\sigma_{nn}^2$.
\end{itemize}
\end{minipage}
\end{algorithm}

\begin{remark}
    We note that variational temporal discretization of the energy-dissipation law, such as the Jordan-Kinderleherer-Otto (JKO) type scheme \cite{JKOscheme} can be used to formulate the loss function in learning problems. For instance, see the paper \cite{JKONet} and the references therein. Compared with the JKO-based approach, our learning framework has advantage of avoiding computing the Wasserstein distance and does not need to solve the forward problem repeatedly for matching data.
\end{remark}

\section{Numerical Examples}\label{sec:examples}
In this section, we will investigate the performance of the learning framework proposed in the previous section using both density data and particle data from SDE simulations under different settings. Furthermore, we will explore the impacts of data quality and quantity on the learning results.

we consider the SDE~\eqref{eqn:generalized_diff}, i.e. $$d\bm X_t=\bm a(\bm X_t)dt+\sigma(\bm X_t)dW_t,$$ where the drift term $\bm a$ satisfies the fluctuation-dissipation theorem $\bm a =-\frac{1}{2}\sigma^2\nabla\psi$ and the stochastic integral is interpreted as backward It\^{o} integral. Our goal is to identify the potential function $\psi$ or the noise intensity $\sigma^2$. The ground truth potential function $\psi$ and the noise intensity $\sigma^2$ will be specified in each example. It should be noted that the learned potential function can be shifted by a constant, as adding a constant to the potential function does not affect the system’s evolution.

In all the examples, we use a constant weighting function $\lambda\equiv 1$ in the loss functions \eqref{eqn:loss_density} and \eqref{eqn:loss_particle2density}. For training, we employ a fully-connected neural network with one hidden layer and 32 nodes per layer to approximate the unknown potential $\psi$ for noise intensity $\sigma^2$. The activation function is {\bf tanh()}, and we use the {\bf Adam} optimizer with an initial learning rate of $5\times 10^{-4}$ in all examples. The learning rate is decayed by a factor of 0.9 every 2,000 epochs. The neural network is trained for 50,000 epochs with the batch size 5 in most of the examples, unless otherwise specified.


\subsection{Learning potential function}
In the first numerical study, we focus on the performance of the density-based and particle-to-density methods for learning the potential function $\psi$ in two different cases.
\begin{example}\label{ex:double-well_potential}
    We consider the potential function $\psi(x)=\frac{1}{2}x^4-x^2$ and the noise intensity $\sigma(x)=\frac{1}{x^2+1}$. Our goal is to identify the potential function $\psi$ with the given noise intensity.

The training data are generated either by simulating the corresponding PDE \eqref{eqn:FPE_generalizedDiff} on the bounded domain $\Omega = [-8, 8]$ using a spatial grid size of $\Delta x = 0.05$ and a time step of $\Delta t = 0.001$, or by estimating the density function $f$ from particle distribution obtained by simulating the SDE \eqref{eqn:generalized_diff}. A total of $M$ initial conditions are used with each initial condition having a Gaussian profile $\mathcal{N}(\mu,0.2^2)$, where the mean value $\mu$ is drawn uniformly from the interval $[-2,2]$. 
    
    We choose the snapshots at $t_1=0.495$, $t=0.5$ and $t_2=0.505$ as our training data and denote the training data by $\{(f_j(x_{i},t_1),f_j(x_{i},t), f_j(x_{i},t_2))\}_{i,j=1}^{N,M}$ (so the observation time step size is $\delta t=5\Delta t$ where $\Delta t$ is the time step used in the PDE or SDE solver). Since the loss function \eqref{eqn:loss_density} is in an integral form, the potential function $\psi$ cannot be uniquely determined using a single group of density data ($M=1$). Therefore, we choose to use multiple groups of data here. Figure~\ref{fig:noiselessData_potential_density-based} shows the learned potential function $\psi_{nn}$ using the density-based method described in Sec.~\ref{sec:density-based} alongside the target $\psi$ for the given $\sigma(x)=\frac{1}{x^2+1}$, with $M=2,5,10,20$ groups of data. As expected, the performance of our method improves as the number of data groups increases. Figure~\ref{fig:noiselessData_potential_particle2density} shows the learned potential function with the same values of $M$ but using the particle-to-density method ($5,000$ particles for each $j$) described in Sec.~\ref{sec:particle-to-density}. For the same value of $M$, the density-based method outperforms the particle-to-density approach that incurs additional approximation error during the density estimation step. Nevertheless, the particle-to-density method still produces satisfactory results and may offer advantages in high-dimensional settings—an aspect we leave for future investigation.

    To further assess the robustness of our method, we examine its performance under varying levels of observation noise. We emphasize that this observation noise in the training data is {\em not} related to the physical noise in the SDE \eqref{eqn:generalized_diff}. For illustration, we focus on the density-based method. The clean training data $\{(f_j(x_{i},t_1),f_j(x_{i},t), f_j(x_{i},t_2))\}_{i,j=1}^{N,M}$ is convoluted with a Gaussian kernel with zero mean and varying standard deviations. In this case, we generate $M=15$ groups of data for training. The resulting learned potential functions $\psi_{nn}$ for different noise levels are shown in Figure~\ref{fig:noisyData_potential_density-based}. 
    
    Next, we introduce a possibly more practical metric to evaluate the learned potential $\psi_{nn}$ in certain real-world applications. We compute the numerical solution to the FPE \eqref{eqn:FPE_generalizedDiff} with the learned potential $\psi_{nn}$, denoted by $f_{nn}$, and compare with the true density $f$ by measuring the relative difference in $L_2$-norm, i.e.,
    \begin{equation}
        d_f (t) := \frac{\| f_{nn} (\cdot, t) - f (\cdot, t) \|_2}{\| f(\cdot, t) \|_2}.
        \label{eqn:forward_error}
    \end{equation}
    Figure~\ref{fig:noisyData_potential_L2error} shows the evolution of the difference $d_f(t)$ from the learned potential functions with different noise levels. Moreover, we simulate the Fokker-Planck equation using the learned potential function and report the relative $L_2$ errors between the learned and true densities at various time points in Figure~\ref{fig:noisyData_potential_L2error}, as this may serve as a more practical metric for certain real-world applications. Moreover, in Figures~\ref{fig:noisyData_potential_forwardComparasion_noiselevel=0.0}, \ref{fig:noisyData_potential_forwardComparasion_noiselevel=0.6}, and~\ref{fig:noisyData_potential_forwardComparasion_true}, we present the numerical solutions to the Fokker-Planck equation \eqref{eqn:FPE_generalizedDiff} with the potential $\psi$ replaced by the learned potential $\psi_{nn}$ and the density $f$ corresponding to the ground truth $\psi$. For simplicity, we only show the learned solutions using clean training data and noisy training data with a noise level of $0.6$ in Figure~\ref{fig:noisyData_potential_forwardComparasion_noiselevel=0.0} and Figure~\ref{fig:noisyData_potential_forwardComparasion_noiselevel=0.6}.
    
    \begin{figure}[htbp]
    \centering
    \begin{subfigure}[b]{0.47\textwidth}
        \centering
        \includegraphics[width=0.95\textwidth]{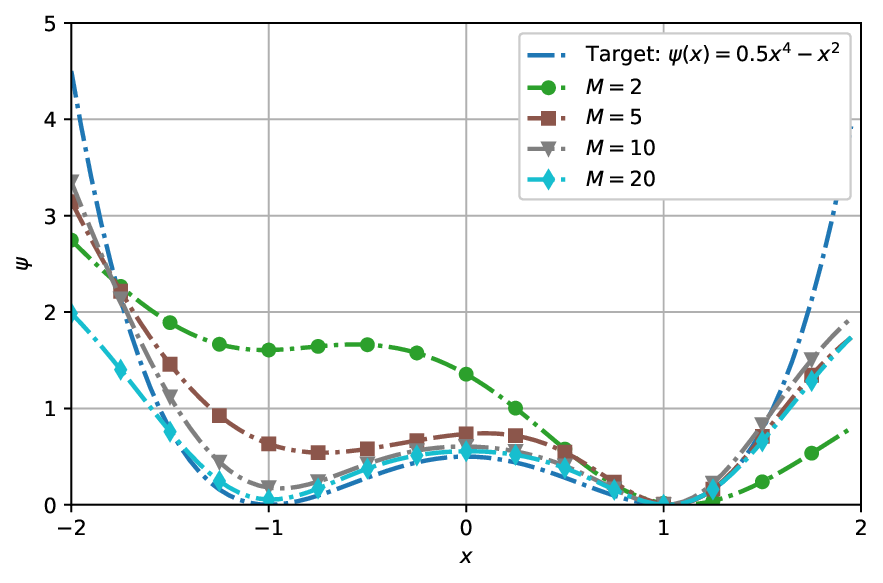}
        \caption{Potential $\psi$, density-based}
        \label{fig:noiselessData_potential_density-based}
    \end{subfigure}
    \hfill
    \begin{subfigure}[b]{0.47\textwidth}
        \centering
        \includegraphics[width=0.95\textwidth]{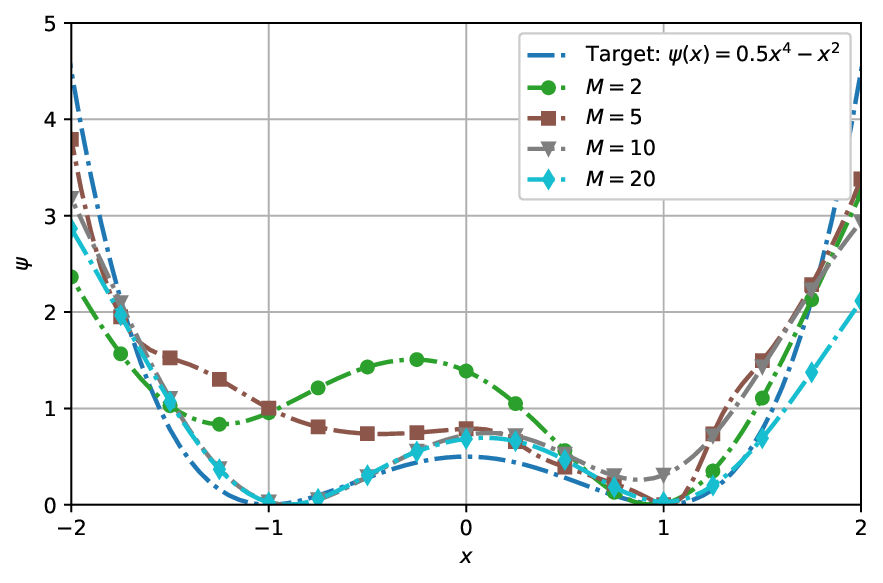}
        \caption{Potential $\psi$, particle-to-density}
        \label{fig:noiselessData_potential_particle2density}
    \end{subfigure}
    \caption{\it  The learned potential function $\psi_{nn}$ resulting from different number $M$ of groups of training data sets compared with the ground truth $\psi = 0.5 x^4 - x^2$  in the one-dimensional case of \eqref{eqn:generalized_diff} with given noise intensity $\sigma(x)=\frac{1}{x^2+1}$.  (a) Using the density-based method;   (b) Using the  particle-to-density method. }
    \label{fig:noiselessData_potential}
    \end{figure}

\begin{figure}[htbp]
    \centering

    \begin{subfigure}[b]{0.47\textwidth}
        \centering
        \includegraphics[width=0.95\textwidth]{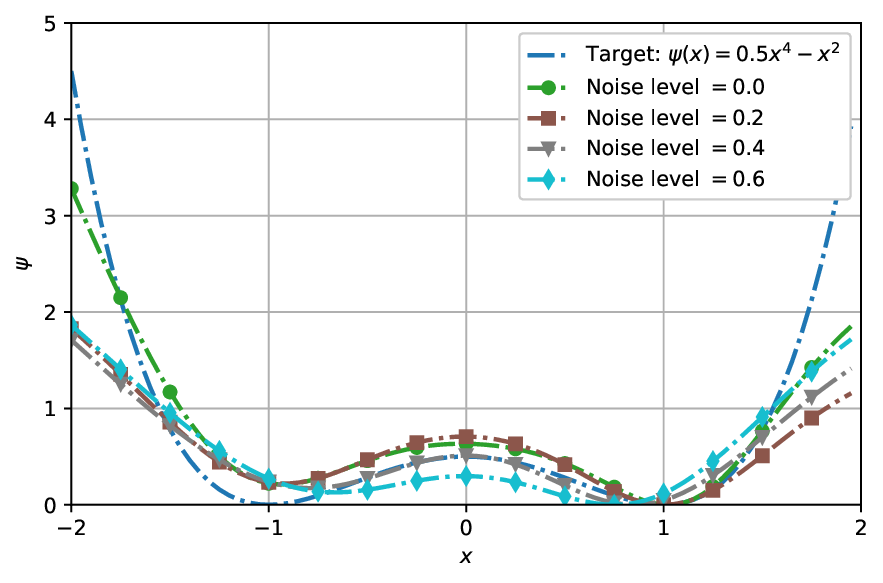}
        \caption{potential, noisy data}
        \label{fig:noisyData_potential_density-based}
    \end{subfigure}
    \hfill
    \begin{subfigure}[b]{0.47\textwidth}
        \centering
        \includegraphics[width=0.95\textwidth]{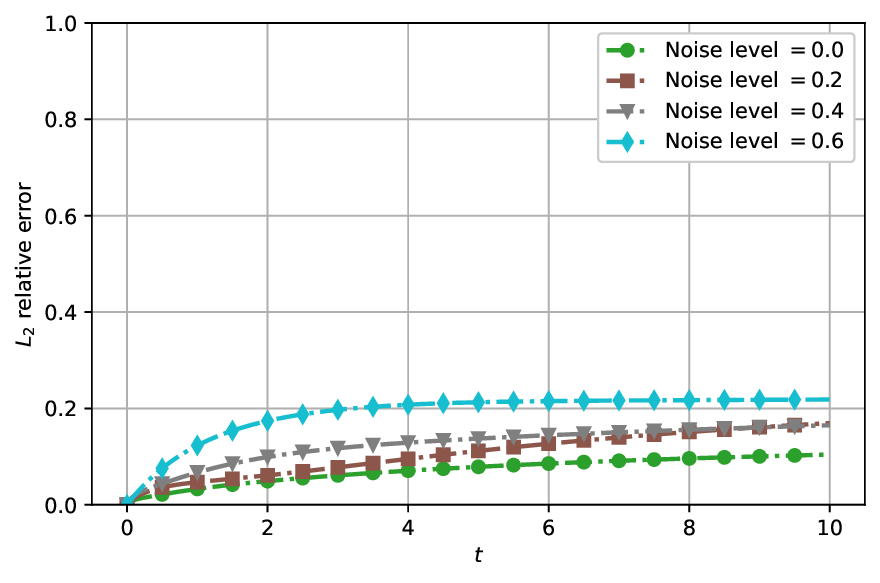}
        \caption{relative $L_2$ errors}
        \label{fig:noisyData_potential_L2error}
    \end{subfigure}
    \\
    \begin{subfigure}[b]{0.32\textwidth}
        \centering
        \includegraphics[width=0.95\textwidth]{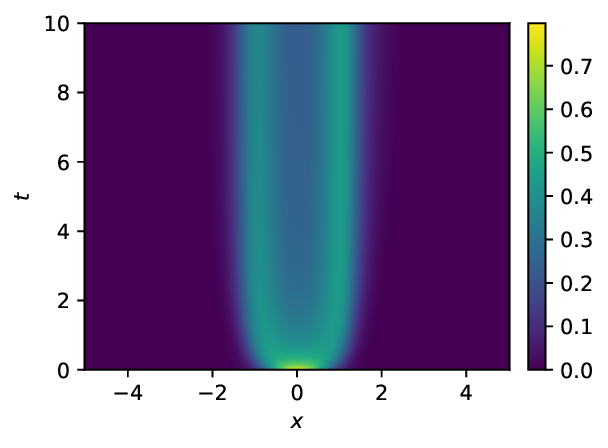}
        \caption{Noise level $=0.0$}
        \label{fig:noisyData_potential_forwardComparasion_noiselevel=0.0}
    \end{subfigure}
    \hfill
    \begin{subfigure}[b]{0.32\textwidth}
        \centering
        \includegraphics[width=0.95\textwidth]{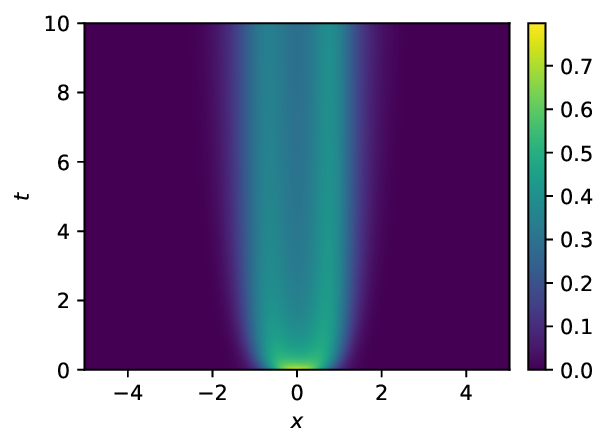}
        \caption{Noise level $=0.6$}
        \label{fig:noisyData_potential_forwardComparasion_noiselevel=0.6}
    \end{subfigure}
    \hfill
    \begin{subfigure}[b]{0.32\textwidth}
        \centering
        \includegraphics[width=0.95\textwidth]{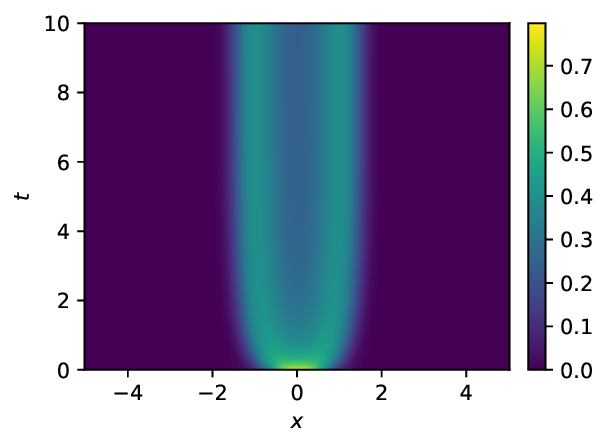}
        \caption{Ground truth}
        \label{fig:noisyData_potential_forwardComparasion_true}
    \end{subfigure}
    \caption{\it  (a) The learned potential function $\psi_{nn}$ resulting from training with different levels of noise and fixed number of groups of data ($M=15$) compared with the ground-truth potential $\psi$ and using the density-based method. (b) The relative $L_2$ difference $d_f(t)$ \eqref{eqn:forward_error} of the forward solutions to  the Fokker-Planck equation~\eqref{eqn:FPE_generalizedDiff} using the learned potentials $\psi_{nn}$ and the ground truth $\psi$. The solutions of the Fokker-Planck equation~\eqref{eqn:FPE_generalizedDiff} using the learned potentials $\psi_{nn}$ with noise level $=0.0$ training data (c), noise level $=0.6$ (d) and the ground truth (e).}
    \label{fig:noisyData_potential}
    \end{figure}

\end{example}

\begin{example}
    In the second example, we intend to illustrate the ill-posedness of learning the potential function $\psi$ from the training data and its relationship to the properties of the training data. Let's revisit the energy-dissipation law~\eqref{eqn:EnergyLaw_generalizedDiff} as follows
    \begin{equation}
        \frac{dE}{dt}=-\int_\Omega\frac{f}{\sigma^2/2}|\bm u|^2d\bm x,
    \end{equation}
    where the free energy and the velocity $\bm u$ are defined by 
    \begin{equation} \label{eqn:fenvel}
        E[f] = \int_\Omega \left[f\ln f+\psi f\right]d\bm x, \qquad \bm u=-\left(\frac{\sigma^2}{2}\nabla\ln f+\frac{\sigma^2}{2}\nabla\psi\right).
    \end{equation}
    The unknown function $\psi$ appears on both sides of the energy-dissipation law, leading to an inverse problem that is generally ill-posed, as one seeks to recover the potential function $\psi$ from the integral and the nonconvex loss function. This motivates us, in Example~\ref{ex:double-well_potential}, to select $M$ groups of initial data as Gaussian-type test functions trying to better determine the gradient of the potential function. However, the ill-posed problem can be avoided by using steady-state data. In the steady state, the time derivative of the energy equals zero, i.e., $\frac{dE}{dt}=0$. Moreover, the right-hand side of the energy-dissipation law reaches its unique minimizer when the velocity $\bm u=0$. Noting that the noise intensity $\sigma^2$ is specified and nonzero, it follows from the expression of $\ub$ in \eqref{eqn:fenvel} that the gradient of the potential function is uniquely determined by the density function $f$ corresponding to the training data.
    
    To illustrate this observation, we aim to learn a triple-well potential $\psi$ using the training data at different time instances, with the noise intensity $\sigma^2(x)=\left[1+\frac{1}{2}\cos(3x+\frac{1}{2})\right]^2$ provided. The training data are obtained by solving the Fokker-Planck equation \eqref{eqn:FPE_generalizedDiff} using a similar setting of the previous example and the observation time step size is still chosen as $\delta t = 5\Delta t$. Figure~\ref{fig:energy} shows the evolution of the free energy $E$. Figure~\ref{fig:unsteady-steady_potential} shows the learned triple-well potentials using one group ($M=1$) of training data at time $t=20$ (unsteady state in this case) and using one group at time $t=200$ (steady state) compared with the ground truth.  The results indicate that the triple-well potential can be learned from either set of training data, but the latter is more accurate than the former because it avoids the ill-posedness of the problem.

\begin{figure}[htbp]
    \centering
    \begin{subfigure}[b]{0.47\textwidth}
        \centering
        \includegraphics[width=0.95\textwidth]{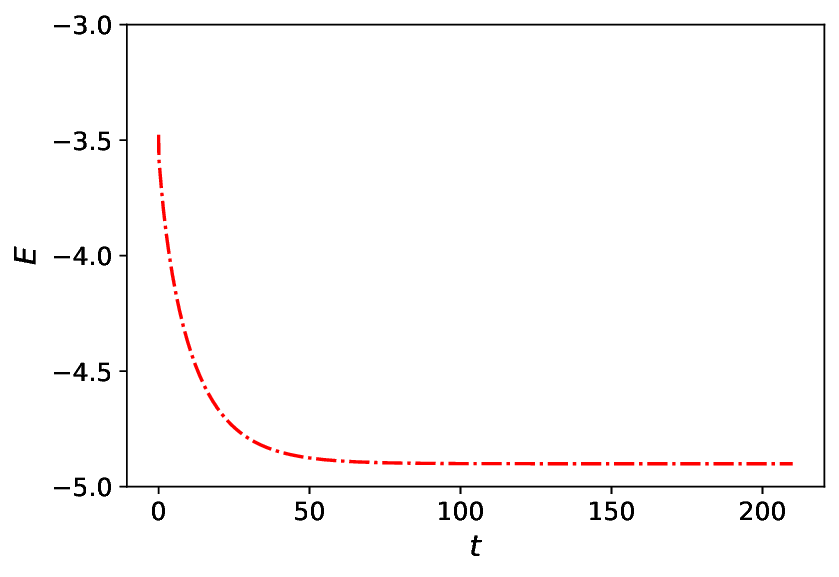}
        \caption{The evolution of the energy}
        \label{fig:energy}
    \end{subfigure}
    \hfill
    \begin{subfigure}[b]{0.47\textwidth}
        \centering
        \includegraphics[width=0.95\textwidth]{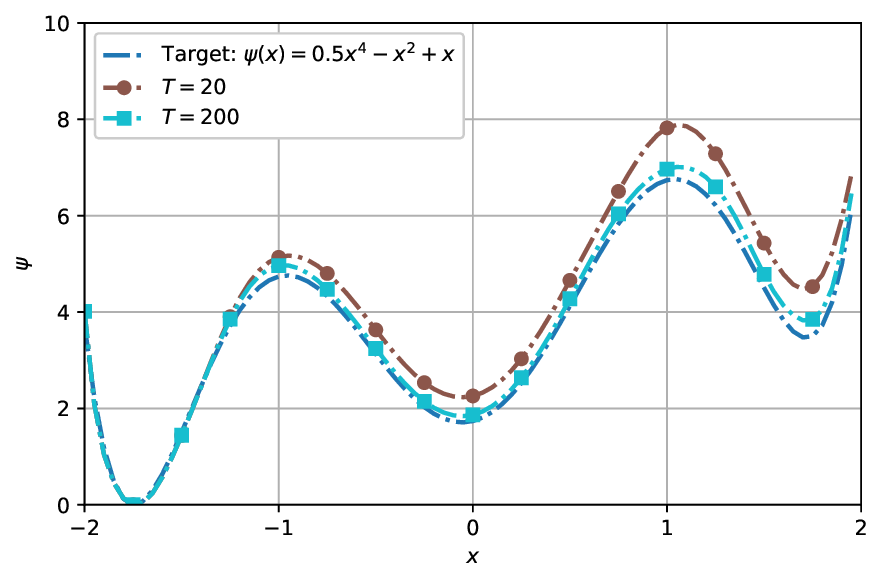}
        \caption{Potential}
        \label{fig:unsteady-steady_potential}
    \end{subfigure}
    \caption{\it (a) The evolution of the energy $E(t)$ for the case with $\psi= \frac{1}{2} x^4 - x^2 +x$ and $\sigma^2=(1+\cos(3x+\frac{1}{2}))^2$. (b) The learned potential functions $\psi$ using one group of training data ($M=1$) at an unsteady state ($t=20$) and at the steady state ($t=200$) compared with the ground truth potential $\psi$.}
    \label{fig:unsteady-steady}
    \end{figure}

\end{example}

\subsection{Learning noise intensity $\sigma$}
In this section, we evaluate the performance of the density-based  method (Sec.~\ref{sec:density-based} and the particle-to-density method (Sec.~\ref{sec:particle-to-density}) for learning the noise intensity $\sigma^2$.
\begin{example}\label{ex:double-well_noise}
    We again consider the case with the potential function $\psi(x)=\frac{1}{2}x^4-x^2$ and the noise intensity $\sigma(x)=\frac{1}{x^2+1}$ and aim to learn the noise intensity $\sigma^2$ with the given potential function $\psi$. The training data $\{(f_j(x_{i},t_1),f_j(x_{i},t), f_j(x_{i},t_2))\}_{i,j=1}^{N,M}$ are obtained by solving the Fokker--Planck equation \eqref{eqn:FPE_generalizedDiff} in a bounded domain $\Omega=[-8,8]$ with grid size $\Delta x=0.05$ and time step size $\Delta t=0.001$ or estimating the density function $f$ from the SDE \eqref{eqn:generalized_diff} particles. We simulate $M$ different initial distributions of $\mathcal{N}(\mu,0.2^2)$, where the mean values $\mu$ are uniformly spaced in domain $[-2,2]$, and choose the snapshots at $t_1=0.495$, $t=0.5$ and $t_2=0.505$ as our training data (so the observation time step size is $\delta t=5\Delta t$). Figure~\ref{fig:noiselessData_sigma_density-based} shows the learned noise intensity $\sigma_{nn}^2$ for the given potential $\psi(x)=\frac{1}{2}x^4-x^2$ along with the target $\sigma^2(x)=\frac{1}{(x^2+1)^2}$ using the density-based method. Figure~\ref{fig:noiselessData_sigma_particle2density} shows the learned noise intensity using the particle-to-density method ($10,000$ particles for each $j$). As in Example~\ref{ex:double-well_potential}, the density-based method outperforms the particle-to-density method. The particle-to-density method still provides a reasonable profile of the noise intensity. It can be observed that the learned noise intensity $\sigma^2$ in Figure~\ref{fig:noiselessData_sigma} is more accurate than the learned potential function $\psi$ in Figure~\ref{fig:noiselessData_potential}. This suggests that our method may be less robust in learning the potential function $\psi$ compared to the noise intensity $\sigma^2$. This difference is not coincidental and could be attributed to the following two main facts. First, the unknown $\sigma^2$ {\em only} appears in the dissipation rate of the energy-dissipation law (i.e. the second term in \eqref{eqn:loss_density}) and is absent in the first term of \eqref{eqn:loss_density}), which renders the loss function being convex with respect to $\sigma^2$. Second, training data are sampled from the initial stage of the system evolution using $M$ different initial conditions uniformly distributed in the domain $[-2, 2]$, ensuring that the data have sufficient spatial coverage. 


    \begin{figure}[htbp]
    \centering
    \begin{subfigure}[b]{0.47\textwidth}
        \centering
        \includegraphics[width=0.95\textwidth]{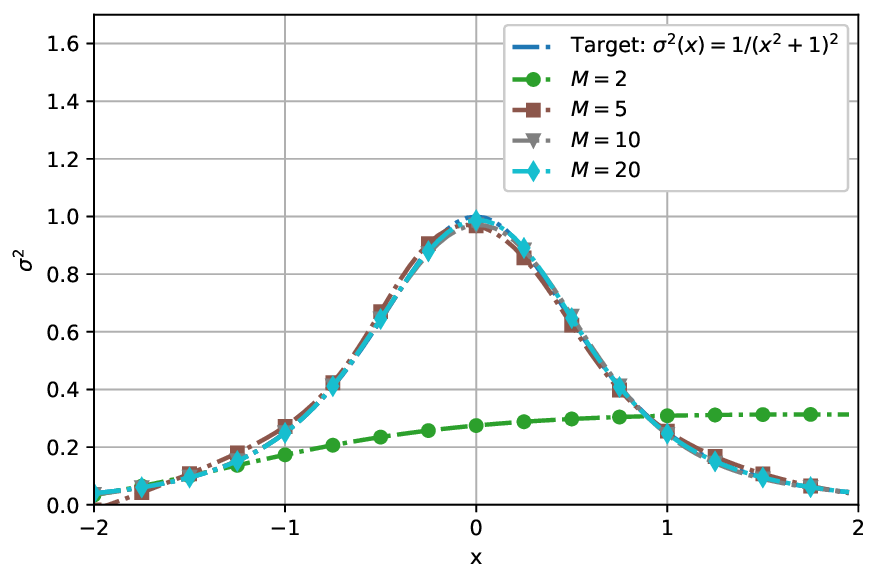}
        \caption{Noise intensity $\sigma^2$, density-based}
        \label{fig:noiselessData_sigma_density-based}
    \end{subfigure}
    \hfill
    \begin{subfigure}[b]{0.47\textwidth}
        \centering
        \includegraphics[width=0.95\textwidth]{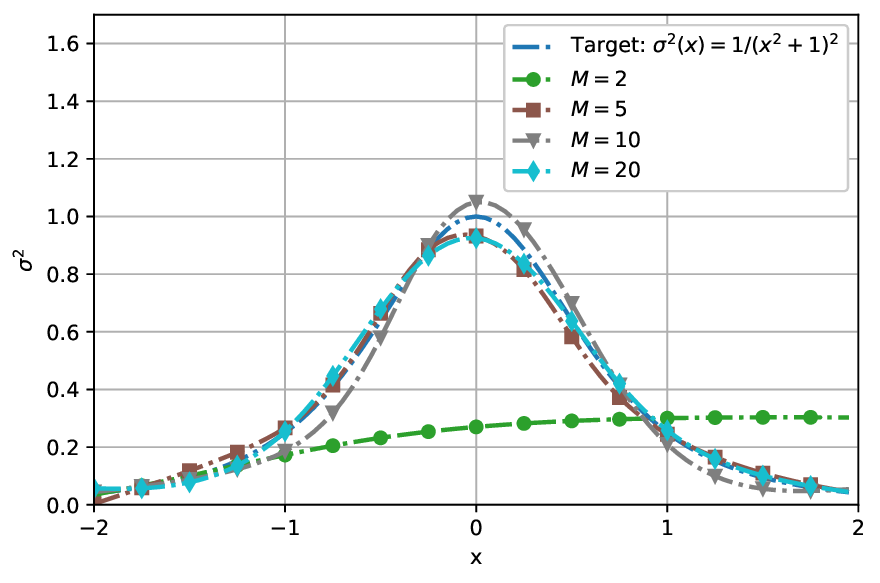}
        \caption{Noise intensity $\sigma^2$, particle-to-density}
        \label{fig:noiselessData_sigma_particle2density}
    \end{subfigure}
    \caption{\it  The learned noise intensity $\sigma_{nn}^2$ resulting from different numbers of groups of training data sets $M$, compared with the ground truth $\sigma^2(x) = \frac{1}{(1+x^2)^2}$ in the one-dimensional case \eqref{eqn:generalized_diff} with the given potential $\psi(x) = \frac{1}{2} x^4 - x^2$.  (a) Using the density-based method; (b)  Using the particle-to-density method.}
    \label{fig:noiselessData_sigma}
    \end{figure}
    
\end{example}

\subsection{Corrupted observations}
\begin{example}\label{ex:PINN}
    {\bf (Corrupted observations, EnVarA vs PDE-based method)}. In this example, we provide a simple comparison between our EnVarA-based learning framework and a PDE-based learning framework for corrupted observations aiming to show the robustness of our method. To be more specific, motivated by the PDE-based learning framework \cite{PINNinverse1,Yannis1,Yannis2,Yannis3}, we construct the following loss function based on the Fokker--Planck equation \eqref{eqn:FPE_generalizedDiff}
    \begin{align}
        {\rm L}_{\rm PDE}=\frac{1}{NM}\sum\limits_{i,j=1}^{N,M}\left[\frac{f_{i,j}(t_2)-f_{i,j}(t_1)}{2\delta t}+\grad\cdot\left(\bm a_i f_{i,j}(t)\right)-\frac{1}{2}\grad\cdot\left(\sigma_i^2\grad f_{i,j}(t)\right)\right]^2,
    \end{align}
    where the drift term $\bm a=-\frac{\sigma^2}{2}\grad\psi$, $f_{i,j}(t)=f_j(x_i,t)$, $\bm a_i=\bm a(x_i)$, and $\sigma_i=\sigma(x_i)$. The spatial derivatives are discretized using the central difference scheme instead of automatic differentiation. In practice, the potential function $\psi$ or the noise intensity $\sigma$ should be replaced by a neural network. We choose a non-symmetric double-well potential function $\psi=\frac{1}{2}x^4-x^2+x$ and a constant noise intensity $\sigma=1.5$ as the ground truths. For simplicity, we assume the noise intensity is known and aim to learn the potential function from the steady-state density data.  
    
    The training data $\{(f_j(x_{i},t_1),f_j(x_{i},t), f_j(x_{i},t_2))\}_{i,j=1}^{N,M}$ are obtained by solving the Fokker--Planck equation \eqref{eqn:FPE_generalizedDiff} in a bounded domain $\Omega=[-8,8]$ with grid size $\Delta x=0.05$ and time step size $\Delta t=0.001$ or estimating the density function $f$ from the SDE \eqref{eqn:generalized_diff} particles. We select only one initial profile $\mathcal{N}(\mu,0.2^2)$ with the mean value $\mu$ randomly selected in domain $[-2,2]$ and choose the snapshots at $t_1=199.995$, $t=200$ and $t_2=200.005$ as our training data (so the observation time step size is $\delta t=5\Delta t$), i.e., the hyperparameter $M=1$ in the loss function \eqref{eqn:loss_density}. We artificially destroy the value of the density data at two grid points $x_1$ and $x_2$. Specifically, the density data at $x_1$ is perturbed by adding noise $\alpha\epsilon$ to the raw data ($\Tilde{f}(x_1)=f(x_1)+\alpha\epsilon$), while the density data at $x_2$ is perturbed by subtracting the same value, $\alpha\epsilon$, ($\Tilde{f}(x_2)=f(x_2)-\alpha\epsilon$) to ensure that the integral of the density function remains equal to one. Here, $\alpha$ represents the noise ratio, and $\epsilon$ is the maximum value of the density function over the domain. See Figure~\ref{fig:density_based_potential_noisyObs_PINNvsEnVarA} (a) for the clean and corrupted training data.  In this example, the ratio is selected as $\alpha=0.2$. The learned potential functions using the PDE-based method and the EnVarA-based method with clean training data and corrupted training data are shown in Figure~\ref{fig:density_based_potential_noisyObs_PINNvsEnVarA} (b) and Figure~\ref{fig:density_based_potential_noisyObs_PINNvsEnVarA} (c) respectively.  It is not surprising that our method is more robust than the discrete version of the PDE-based approach, since our EnVarA-based method does not require computing the second derivative of the density function and our loss function is in an integral form.
    
    However, it should be noticed that this is a discrete version of PINN rather than the method proposed in \cite{PINNinverse2, BiLOforPDE} since we did not use automatic differentiation here. Moreover, we employ density data as training data instead of particle data used in \cite{PINNinverse2}, which provides impressive results for learning stochastic differential equation with Brownian motion or L\'evy motion. It is worth mentioning that the methods proposed in \cite{PINNinverse2,BiLOforPDE} may mitigate the impact of corrupted observations, as they defined a more robust loss function. A more comprehensive comparison is left for future work.

    \begin{remark}
        In our setting, the corresponding PDE can be derived from an energy-dissipation law, analogous to the relationship between a primitive function and its derivative in calculus. This inherent structure justifies the design of our loss function, which not only relaxes the regularity requirements for solutions of the Fokker--Planck equation but also imposes minimal smoothness constraints on the unknown potential function itself. Moreover, by circumventing the need for strong regularity conditions on both the solution and the potential function, our method exhibits greater robustness to noisy data compared to traditional PDE-based approaches. However, when the noise level of data is low and the functions are sufficiently smooth, the PDE-based methods are expected to outperform ours, as it employs pointwise loss functions whereas ours relies on an integral form.
    \end{remark}

\begin{figure}[!htb]
    \centering
    \begin{overpic}[width=0.315 \textwidth]{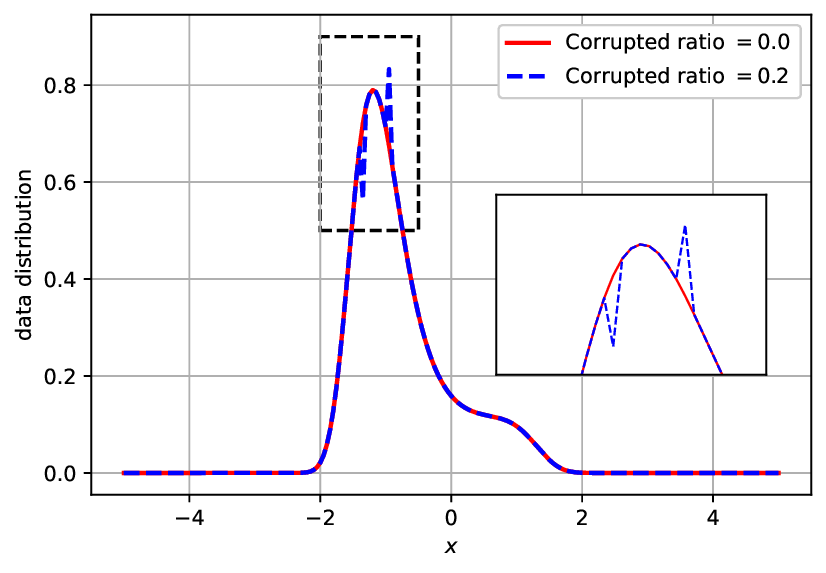}
    \put(0, 64){\scriptsize (a)}
    \end{overpic}
    \hfill
    \begin{overpic}[width=0.334 \textwidth]{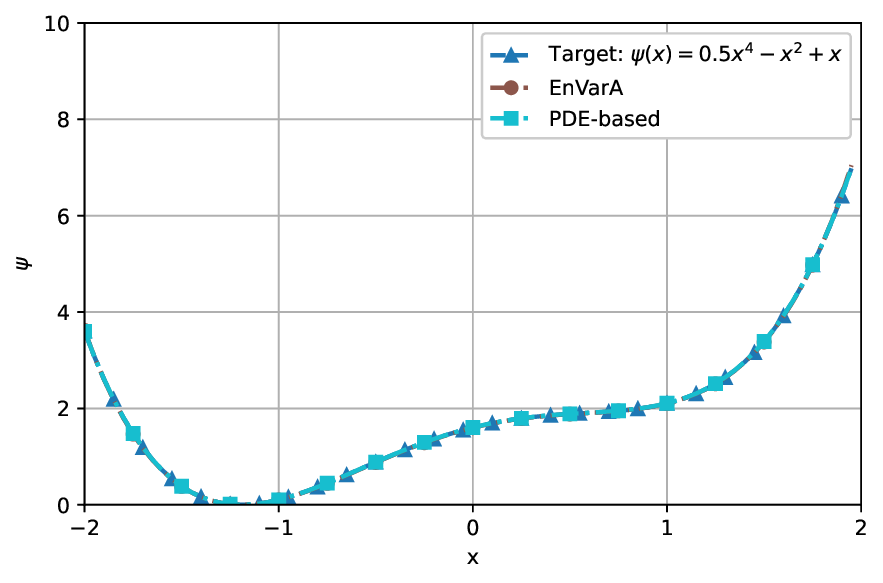}
    \put(-2, 62){\scriptsize (b)}
    \end{overpic}
    \hfill
    \begin{overpic}[width=0.334 \textwidth]{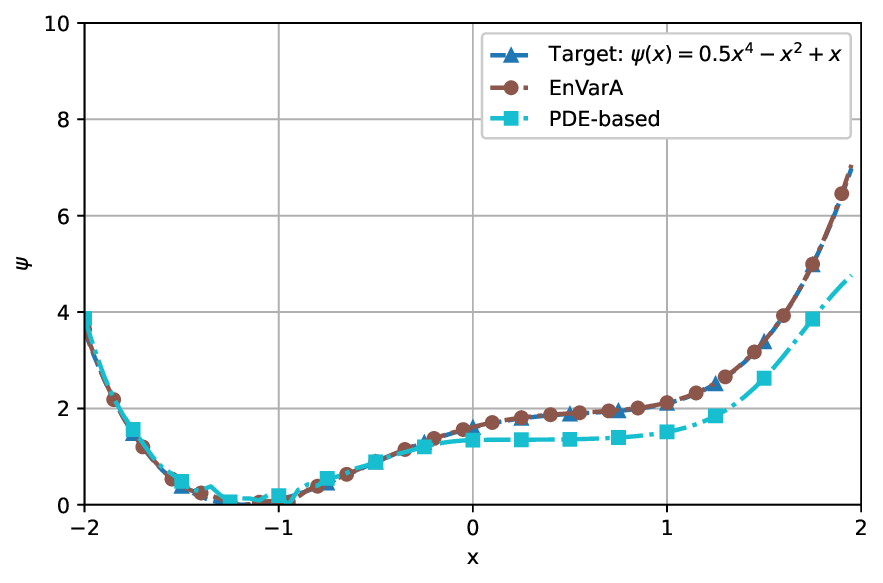}
   \put(-2, 62){\scriptsize (c)}
    \end{overpic}
     \caption{ \it (a) Clean and corrupted training data. (b) The learned potential function $\psi_{nn}$ from the clean training data using our EnVarA-based method and the PDE-based method, compared with the true potential $\psi$. (c) The same as (b) except from the corrupted training data. } 
    \label{fig:density_based_potential_noisyObs_PINNvsEnVarA}
\end{figure}

\end{example}

\subsection{A 2D example}
In this section, we examine the particle-to-density method in a two-dimensional (2D) system~\eqref{eqn:generalized_diff}. 

\begin{example} 
We consider the system with the potential function $\psi(x,y)=\frac{1}{4}x^4-\frac{1}{2}x^2+\frac{1}{4}y^4-\frac{1}{2}y^2$ and the noise intensity $\sigma=\sqrt{2}$, and aim to learn the potential function $\psi$ with the given noise intensity $\sigma^2$. The training data are obtained by solving the SDE \eqref{eqn:generalized_diff} with time-step size $\Delta t=0.001$.  $M$ groups of training data are generated from $M$ different initial conditions of the profile $\mathcal{N}(\mu,2I)$, where the mean values $\mu$ are uniformly randomly distributed in the domain $[-1.5,1.5]\times[-1.5,1.5]$ and $I$ is the $2\times 2$ identity matrix. We choose the snapshots at $t_1=1.798$, $t=1.799$ and $t_2=1.8$ as our training data (so the observation time step size is $\delta t=\Delta t$). The grid sizes for evaluating integrals is chosen as $\Delta x=\Delta y=0.1$. The activation function is {\bf tanh()}, and we use the {\bf Adam} optimizer with an initial learning rate of $5\times 10^{-4}$. The learning rate is decayed by a factor of $0.9$ every $2,000$ epochs. The neural network is trained for $20,000$ epochs with the batch size 5.


\begin{figure}[!htbp]
    \centering
    \includegraphics[width=0.75\textwidth]{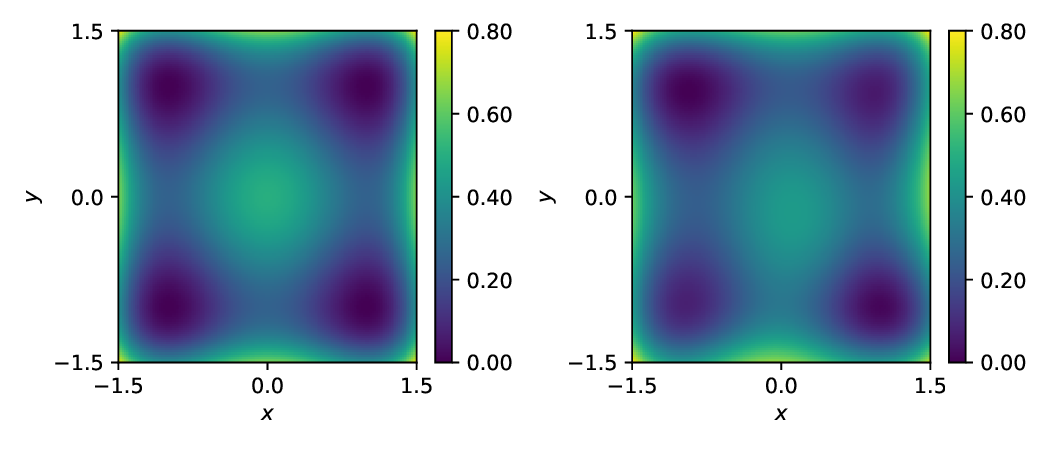}
    \caption{\it Learning the 2D target potential function $\psi(x,y)=\frac{1}{4}x^4-\frac{1}{2}x^2+\frac{1}{4}y^4-\frac{1}{2}y^2$ using the particle-to-density method. The density plot of the  ground truth $\psi$ (left) and the learned potential function $\psi_{nn}$ (right).} 
    \label{fig:particle2density_potential_2dEx_2ddisplay}
    \end{figure}

Figure~\ref{fig:particle2density_potential_2dEx_2ddisplay} compares the density plot of the learned potential $\psi_{nn}$ resulting from $30$ groups of clean training data ($M=30$) with that of the ground truth $\psi(x,y)$. In this scenario, each density function $f$ is estimated using $10,000$ particles. The profile of the learned potential appears to be close to that of the true potential as shown in Figure~\ref{fig:particle2density_potential_2dEx_2ddisplay}.
Furthermore, similar to Example~\ref{ex:double-well_potential}, we assess the learned potential $\psi_{nn}$ by computing the relative difference $d_f$ defined in~\eqref{eqn:forward_error} between the predicted density function $f_{nn}$ and the true $f$ obtained by simulating the SDE~\eqref{eqn:generalized_diff} with the learned potential $\psi_{nn}$ and the true potential $\psi$ respectively. Figure~\ref{fig:particle2density_potential_2dEx_L2error_30groups} shows the difference $d_f$ suggesting that the learned potential provides reasonable results in terms of the practical metric $d_f$.

To further validate our method in the 2D setting, we examine its performance by adding noise to the training data as in Example~\ref{ex:double-well_potential} and varying the number of particles $N$ in \eqref{eqn:loss_particle2density} and \eqref{eqn:discrete_energy_Riemann_particle2density}. As in the 1D case, we obtain the noisy data by convoluting the clean training data with a zero-mean Gaussian kernel with covariance $0.04 I$. In this case, we use $50$ groups of noisy training data ($M=50)$. Figure~\ref{fig:particle2density_potential_2dEx_diffNparticles} compares the learned potentials obtained with the noisy data and different number of particles $N$ against the ground truth. In addition,  Figure~\ref{fig:particle2density_potential_2dEx_L2error_Nparticles} evaluates the learned potentials $\psi_{nn}$ by presenting the relative difference $d_f$ in \eqref{eqn:forward_error}.  As shown in Figures~\ref{fig:particle2density_potential_2dEx_diffNparticles} and~\ref{fig:particle2density_potential_2dEx_L2error_Nparticles}, the learning results appear reasonable and the accuracy improves as one increases the number of particles $N$. 

 \begin{figure}[!htb]
    \centering
    \begin{subfigure}[b]{0.45\textwidth}
        \centering
        \includegraphics[width=0.9\textwidth]{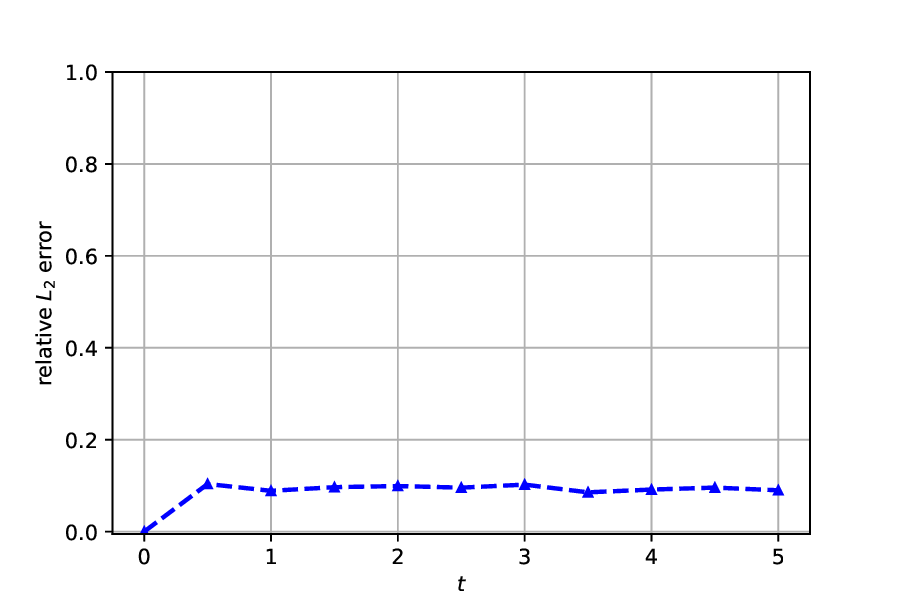}
        \caption{Noise level $=0.0$, $M=30$, $N=1e4$}
        \label{fig:particle2density_potential_2dEx_L2error_30groups}
    \end{subfigure}
    \hfill
    \begin{subfigure}[b]{0.45\textwidth}
        \centering
        \includegraphics[width=0.9\textwidth]{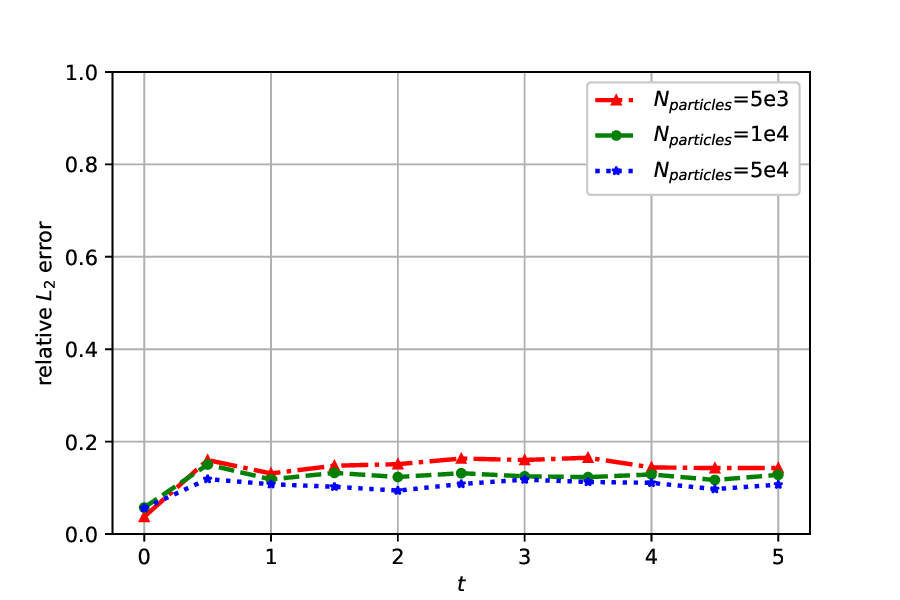}
        \caption{Noise level $=0.2$, $M=15$, $N=5e3, 1e4, 5e4$}
        \label{fig:particle2density_potential_2dEx_L2error_Nparticles}
    \end{subfigure}
    \caption{\it (a) The relative $L_2$ difference $d_f$ in~\eqref{eqn:forward_error} between the true density $f$ and the forward solution $f_{nn}$ resulting from the learned potential $\psi_{nn}$ in Figure~\ref{fig:particle2density_potential_2dEx_2ddisplay} corresponding to the clean training data and $N=10,000$.  (b) The difference $d_f$ between the true $f$ and the forward solution $f_{nn}$ resulting from the learned potentials $\psi_{nn}$ shown in Figure~\ref{fig:particle2density_potential_2dEx_diffNparticles} corresponding to the noisy training data and different number of particles $N=5,000, 10,000$ and $50,000$. } 
    \label{fig:particle2density_potential_2dEx_L2error}
    \end{figure}

  \begin{figure}[!htbp]
    \centering
        \includegraphics[width=0.75\textwidth]{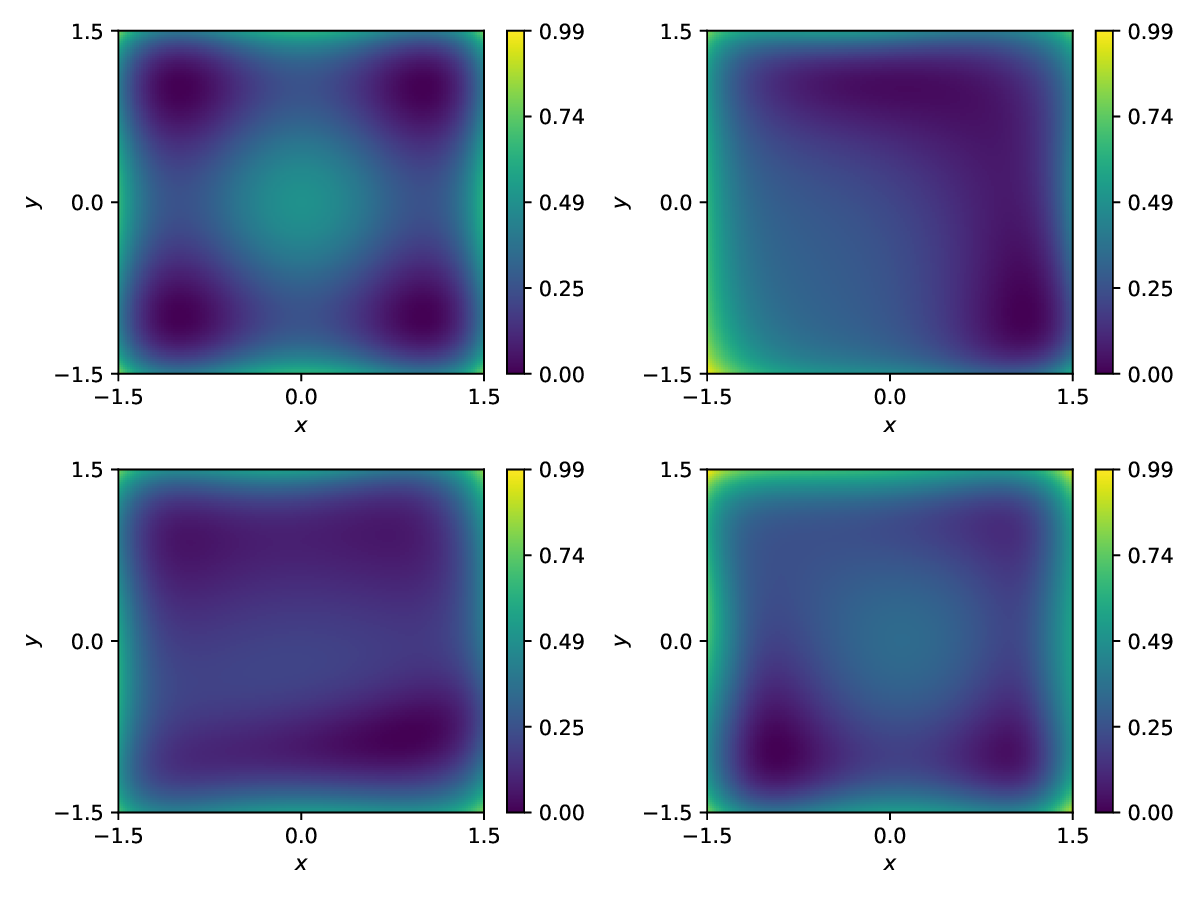}
    \caption{\it The density plots of the 2D ground truth potential $\psi(x,y)=\frac{1}{4}x^4-\frac{1}{2}x^2+\frac{1}{4}y^4-\frac{1}{2}y^2$ (top left) and the learned potential $\psi_{nn}$ from the noisy training data and using $N=5,000$ particles (top right), $10,000$ particles (bottom left), or $50,000$ particles (bottom right).} 
    \label{fig:particle2density_potential_2dEx_diffNparticles}
    \end{figure}

\begin{remark}
    The density-based method is expected to perform better, as it avoids the estimation error associated with reconstructing the density from particles.
\end{remark}

\end{example}

\section{Conclusion}\label{sec:discussion}
We have utilized the energy-dissipation law of the underlying physical systems to derive the new loss function for learning generalized diffusions that accommodate different types of training data (density or particle data). We validated the performance of the proposed methods through several representative examples and investigated the impact of data quality and data property on these methods. Broadly speaking, our approaches offer several advantages, including robustness to corrupted/noisy observations due to the weak-form of the loss function, easy extension to more general physical systems through the widely used energetic variational approach, and potential to handle higher-dimensional challenges. 

One important challenge in our proposed method is handling high-dimensional problems, as density data are not generally readily available. Instead, the density must first be approximated by particle sampling. However, estimating a high-dimensional density function is a key problem in the statistical community. Further investigation is needed to develop a more suitable loss function based directly on particle data, rather than relying on an estimated density function. On the other hand, it is worth noting that loss functions formulated in the weak (variational) form are generally more robust to corrupted or noisy observations than those in the strong form. However, the weak form often struggles to uniquely capture local information. This suggests that combining the two approaches may yield a more effective loss function for learning either the solution of a PDE in forward problems or the coefficients of a PDE in inverse problems. These issues will be investigated in future work.

\section*{Acknowledgments}
Y.~Lu and X.~Li are partially supported by DOE DE-SC0022276. C.~Liu is partially supported by NSF DMS-2216926 and DMS-2410742.
Q.~Tang was partially supported by the U.S. Department of Energy Advanced Scientific Computing Research (ASCR) under DE-FOA-2493 “data-intensive scientific machine learning” and the ASCR program of Mathematical Multifaceted Integrated Capability Center (MMICC). Y.~Wang is partially supported by NSF DMS-2153029 and DMS-2410740.

\bibliographystyle{abbrv}

\begin{thebibliography}{10}

\bibitem{Yannis2}
J.~Anderson, I.~Kevrekidis, and R.~Rico-Martinez.
\newblock A comparison of recurrent training algorithms for time series analysis and system identification.
\newblock {\em Computers \& Chemical Engineering}, 20:S751--S756, 1996.

\bibitem{arnol2013mathematical}
V.~I. Arnol'd.
\newblock {\em Mathematical methods of classical mechanics}, volume~60.
\newblock Springer Science \& Business Media, 2013.

\bibitem{GP4PDEs2}
P.~Batlle, Y.~Chen, B.~Hosseini, H.~Owhadi, and A.~M. Stuart.
\newblock Error analysis of kernel/gp methods for nonlinear and parametric pdes.
\newblock {\em Journal of Computational Physics}, 520:113488, 2025.

\bibitem{Hamiltonian2}
T.~Bertalan, F.~Dietrich, I.~Mezić, and I.~Kevrekidis.
\newblock {On learning Hamiltonian systems from data}.
\newblock {\em Chaos: An Interdisciplinary Journal of Nonlinear Science}, 29(12):121107, 2019.

\bibitem{bongard2007automated}
J.~Bongard and H.~Lipson.
\newblock Automated reverse engineering of nonlinear dynamical systems.
\newblock {\em Proceedings of the National Academy of Sciences}, 104(24):9943--9948, 2007.

\bibitem{SINDy}
S.~Brunton, J.~Proctor, and J.~Kutz.
\newblock Discovering governing equations from data by sparse identification of nonlinear dynamical systems.
\newblock {\em Proceedings of the National Academy of Sciences}, 113(15):3932--3937, 2016.

\bibitem{Koopman1}
M.~Budišić, R.~Mohr, and I.~Mezić.
\newblock {Applied Koopmanisma}.
\newblock {\em Chaos: An Interdisciplinary Journal of Nonlinear Science}, 22(4):047510, 2012.

\bibitem{JKONet}
C.~Bunne, L.~Papaxanthos, A.~Krause, and M.~Cuturi.
\newblock Proximal optimal transport modeling of population dynamics.
\newblock In G.~Camps-Valls, F.~J.~R. Ruiz, and I.~Valera, editors, {\em Proceedings of The 25th International Conference on Artificial Intelligence and Statistics}, volume 151 of {\em Proceedings of Machine Learning Research}, pages 6511--6528. PMLR, 28--30 Mar 2022.

\bibitem{HenonMaps}
J.~W. Burby, Q.~Tang, and R.~Maulik.
\newblock Fast neural poincaré maps for toroidal magnetic fields.
\newblock {\em Plasma Physics and Controlled Fusion}, 63(2), 12 2020.

\bibitem{MTao}
R.~Chen and M.~Tao.
\newblock Data-driven prediction of general {Hamiltonian} dynamics via learning exactly-symplectic maps.
\newblock In {\em Proceedings of the 38th International Conference on Machine Learning}, 2021.

\bibitem{PINNinverse2}
X.~Chen, L.~Yang, J.~Duan, and G.~E. Karniadakis.
\newblock Solving inverse stochastic problems from discrete particle observations using the fokker--planck equation and physics-informed neural networks.
\newblock {\em SIAM Journal on Scientific Computing}, 43(3):B811--B830, 2021.

\bibitem{GP4PDEs1}
Y.~Chen, B.~Hosseini, H.~Owhadi, and A.~M. Stuart.
\newblock Solving and learning nonlinear pdes with gaussian processes.
\newblock {\em Journal of Computational Physics}, 447:110668, 2021.

\bibitem{flowMap1}
Y.~Chen and D.~Xiu.
\newblock Learning stochastic dynamical system via flow map operator.
\newblock {\em Journal of Computational Physics}, 508:112984, 2024.

\bibitem{SymplecticRNN}
Z.~Chen, J.~Zhang, M.~Arjovsky, and L.~Bottou.
\newblock Symplectic recurrent neural networks.
\newblock In {\em International Conference on Learning Representations}, 2020.

\bibitem{flowMap2}
V.~Churchill and D.~Xiu.
\newblock Flow map learning for unknown dynamical systems: Overview, implementation, and benchmarks.
\newblock {\em Journal of Machine Learning for Modeling and Computing}, 4(2):173--201, 2023.

\bibitem{weakPINNs}
T.~De~Ryck, S.~Mishra, and R.~Molinaro.
\newblock Weak physics informed neural networks for approximating entropy solutions of hyperbolic conservation laws.
\newblock In {\em Seminar f{\"u}r Angewandte Mathematik, Eidgen{\"o}ssische Technische Hochschule, Z{\"u}rich, Switzerland, Rep}, volume~35, page 2022, 2022.

\bibitem{Felix}
F.~Dietrich, A.~Makeev, G.~Kevrekidis, N.~Evangelou, T.~Bertalan, S.~Reich, and I.~Kevrekidis.
\newblock {Learning effective stochastic differential equations from microscopic simulations: Linking stochastic numerics to deep learning}.
\newblock {\em Chaos: An Interdisciplinary Journal of Nonlinear Science}, 33(2):023121, 2023.

\bibitem{Ding2022}
L.~Ding, W.~Li, S.~Osher, and W.~Yin.
\newblock A mean field game inverse problem.
\newblock {\em Journal of Scientific Computing}, 92(1):7, 2022.

\bibitem{weinan2021applied}
W.~E, T.~Li, and E.~Vanden-Eijnden.
\newblock {\em Applied stochastic analysis}, volume 199.
\newblock American Mathematical Soc., 2021.

\bibitem{PNP}
B.~Eisenberg, Y.~Hyon, and C.~Liu.
\newblock Energy variational analysis of ions in water and channels: Field theory for primitive models of complex ionic fluids.
\newblock {\em The Journal of Chemical Physics}, 133(10):104104, 09 2010.

\bibitem{KernelLearning4}
J.~Feng, C.~Kulick, and S.~Tang.
\newblock Data-driven model selections of second-order particle dynamics via integrating gaussian processes with low-dimensional interacting structures.
\newblock {\em Physica D: Nonlinear Phenomena}, 461:134097, 2024.

\bibitem{Hamiltonian3}
M.~Finzi, K.~Wang, and A.~Wilson.
\newblock Simplifying {Hamiltonian} and {Lagrangian} neural networks via explicit constraints.
\newblock In {\em Advances in Neural Information Processing Systems}, 2020.

\bibitem{LiPNP}
A.~Flavell, M.~Machen, B.~Eisenberg, J.~Kabre, C.~Liu, and X.~Li.
\newblock A conservative finite difference scheme for {Poisson--Nernst--Planck} equations.
\newblock {\em Journal of Computational Electronics}, 13(1):235--249, 2014.

\bibitem{neuralGalerkin}
H.~Gao, M.~J. Zahr, and J.-X. Wang.
\newblock Physics-informed graph neural galerkin networks: A unified framework for solving pde-governed forward and inverse problems.
\newblock {\em Computer Methods in Applied Mechanics and Engineering}, 390:114502, 2022.

\bibitem{Lu_self-test}
Y.~Gao, Q.~Lang, and L.~Fei.
\newblock Self-test loss functions for learning weak-form operators and gradient flows.
\newblock {\em arXiv preprint arXiv:2412.03506}, 2024.

\bibitem{EnVarA1}
M.-H. Giga, A.~Kirshtein, and C.~Liu.
\newblock Variational modeling and complex fluids.
\newblock {\em {Handbook of Mathematical Analysis in Mechanics of Viscous Fluids}}, pages 1--41, 2017.

\bibitem{Yannis3}
R.~González-García, R.~Rico-Martínez, and I.~Kevrekidis.
\newblock Identification of distributed parameter systems: A neural net based approach.
\newblock {\em Computers \& Chemical Engineering}, 22:S965--S968, 1998.

\bibitem{Hamiltonian4}
S.~Greydanus, M.~Dzamba, and J.~Yosinski.
\newblock Hamiltonian neural networks.
\newblock In {\em Advances in Neural Information Processing Systems}, 2019.

\bibitem{metriplectic1}
A.~Gruber, M.~Gunzburger, L.~Ju, and Z.~Wang.
\newblock Energetically consistent model reduction for metriplectic systems.
\newblock {\em Computer Methods in Applied Mechanics and Engineering}, 404:115709, 2023.

\bibitem{MLforIrreversibleProcess2}
A.~Gruber, K.~Lee, H.~Lim, N.~Park, and N.~Trask.
\newblock Efficiently parameterized neural metriplectic systems.
\newblock In {\em The Thirteenth International Conference on Learning Representations}, 2025.

\bibitem{hernandez2021structure}
Q.~Hern{\'a}ndez, A.~Bad{\'\i}as, D.~Gonz{\'a}lez, F.~Chinesta, and E.~Cueto.
\newblock Structure-preserving neural networks.
\newblock {\em Journal of Computational Physics}, 426:109950, 2021.

\bibitem{Hamiltonian5}
J.~Hu, J.-P. Ortega, and D.~Yin.
\newblock A structure-preserving kernel method for learning hamiltonian systems.
\newblock {\em Mathematics of Computation}, 2025.

\bibitem{hu2024energetic}
Z.~Hu, C.~Liu, Y.~Wang, and Z.~Xu.
\newblock Energetic variational neural network discretizations of gradient flows.
\newblock {\em SIAM Journal on Scientific Computing}, 46(4):A2528--A2556, 2024.

\bibitem{stat-PINNs}
S.~Huang, Z.~He, N.~Dirr, J.~Zimmer, and C.~Reina.
\newblock Statistical-physics-informed neural networks (stat-pinns): A machine learning strategy for coarse-graining dissipative dynamics.
\newblock {\em Journal of the Mechanics and Physics of Solids}, page 105908, 2024.

\bibitem{huang2022variational}
S.~Huang, Z.~He, and C.~Reina.
\newblock Variational onsager neural networks (vonns): A thermodynamics-based variational learning strategy for non-equilibrium pdes.
\newblock {\em Journal of the Mechanics and Physics of Solids}, 163:104856, 2022.

\bibitem{EntropyInformed}
Y.~Jiang, W.~Yang, Y.~Zhu, and L.~Hong.
\newblock Entropy structure informed learning for solving inverse problems of differential equations.
\newblock {\em Chaos, Solitons \& Fractals}, 175:114057, 2023.

\bibitem{SympNets}
P.~Jin, Z.~Zhang, A.~Zhu, Y.~Tang, and G.~Karniadakis.
\newblock {SympNets}: Intrinsic structure-preserving symplectic networks for identifying hamiltonian systems.
\newblock {\em Neural Networks}, 132:166--179, 2020.

\bibitem{JKOscheme}
R.~Jordan, D.~Kinderlehrer, and F.~Otto.
\newblock The variational formulation of the fokker--planck equation.
\newblock {\em SIAM journal on mathematical analysis}, 29(1):1--17, 1998.

\bibitem{variationalPINN}
E.~Kharazmi, Z.~Zhang, and G.~Karniadakis.
\newblock {hp-VPINNs}: Variational physics-informed neural networks with domain decomposition.
\newblock {\em Computer Methods in Applied Mechanics and Engineering}, 374:113547, 2021.

\bibitem{Koopman2}
S.~Klus, F.~Nüske, S.~Peitz, J.-H. Niemann, C.~Clementi, and C.~Schütte.
\newblock Data-driven approximation of the koopman generator: Model reduction, system identification, and control.
\newblock {\em Physica D: Nonlinear Phenomena}, 406:132416, 2020.

\bibitem{kubo1966fluctuation}
R.~Kubo.
\newblock The fluctuation-dissipation theorem.
\newblock {\em Reports on progress in physics}, 29(1):255, 1966.

\bibitem{KernelLearning3}
Q.~Lang and F.~Lu.
\newblock Learning interaction kernels in mean-field equations of first-order systems of interacting particles.
\newblock {\em SIAM Journal on Scientific Computing}, 44(1):A260--A285, 2022.

\bibitem{MLforIrreversibleProcess1}
K.~Lee, N.~Trask, and P.~Stinis.
\newblock Machine learning structure preserving brackets for forecasting irreversible processes.
\newblock In {\em Advances in Neural Information Processing Systems}, 2021.

\bibitem{Phase-field_DeepONet}
W.~Li, M.~Z. Bazant, and J.~Zhu.
\newblock Phase-field deeponet: Physics-informed deep operator neural network for fast simulations of pattern formation governed by gradient flows of free-energy functionals.
\newblock {\em Computer Methods in Applied Mechanics and Engineering}, 416:116299, 2023.

\bibitem{LiuPNP}
C.~Liu, C.~Wang, S.~Wise, X.~Yue, and S.~Zhou.
\newblock A second order accurate, positivity preserving numerical method for the {Poisson--Nernst--Planck} system and its convergence analysis.
\newblock {\em Journal of Scientific Computing}, 97(1):23, 2023.

\bibitem{PME}
C.~Liu and Y.~Wang.
\newblock On lagrangian schemes for porous medium type generalized diffusion equations: A discrete energetic variational approach.
\newblock {\em Journal of Computational Physics}, 417:109566, 2020.

\bibitem{flowMap3}
Y.~Liu, Y.~Chen, D.~Xiu, and G.~Zhang.
\newblock A training-free conditional diffusion model for learning stochastic dynamical systems.
\newblock {\em SIAM Journal on Scientific Computing}, 47(5):C1144--C1171, 2025.

\bibitem{KernelLearning2}
F.~Lu, Q.~An, and Y.~Yu.
\newblock Nonparametric learning of kernels in nonlocal operators.
\newblock {\em Journal of Peridynamics and Nonlocal Modeling}, 2023.

\bibitem{KernelLearning6}
F.~Lu, M.~Maggioni, and S.~Tang.
\newblock Learning interaction kernels in stochastic systems of interacting particles from multiple trajectories.
\newblock {\em Foundations of Computational Mathematics}, 22(4):1013--1067, 2022.

\bibitem{KernelLearning1}
F.~Lu, M.~Zhong, S.~Tang, and M.~Maggioni.
\newblock Nonparametric inference of interaction laws in systems of agents from trajectory data.
\newblock {\em Proceedings of the National Academy of Sciences}, 116(29):14424--14433, 2019.

\bibitem{Lu2}
Y.~Lu, R.~Maulik, T.~Gao, F.~Dietrich, I.~Kevrekidis, and J.~Duan.
\newblock {Learning the temporal evolution of multivariate densities via normalizing flows}.
\newblock {\em Chaos: An Interdisciplinary Journal of Nonlinear Science}, 32(3):033121, 2022.

\bibitem{LiuSWasserstein}
S.~Ma, S.~Liu, H.~Zha, and H.~Zhou.
\newblock Learning stochastic behaviour from aggregate data.
\newblock In M.~Meila and T.~Zhang, editors, {\em Proceedings of the 38th International Conference on Machine Learning}, volume 139 of {\em Proceedings of Machine Learning Research}, pages 7258--7267. PMLR, 18--24 Jul 2021.

\bibitem{Hamiltonian1}
M.~Mattheakis, D.~Sondak, A.~Dogra, and P.~Protopapas.
\newblock {Hamiltonian} neural networks for solving equations of motion.
\newblock {\em Physical Review E}, 105:065305, 2022.

\bibitem{weakSINDy4PDE}
D.~Messenger and D.~Bortz.
\newblock Weak {SINDy} for partial differential equations.
\newblock {\em Journal of Computational Physics}, 443:110525, 2021.

\bibitem{weakSINDy}
D.~Messenger and D.~Bortz.
\newblock Weak {SINDy}: Galerkin-based data-driven model selection.
\newblock {\em Multiscale Modeling \& Simulation}, 19(3):1474--1497, 2021.

\bibitem{weakSINDy4meanfield}
D.~Messenger and D.~Bortz.
\newblock Learning mean-field equations from particle data using {WSINDy}.
\newblock {\em Physica D: Nonlinear Phenomena}, 439:133406, 2022.

\bibitem{weakSINDy4Hamiltonian}
D.~Messenger, J.~Burby, and D.~Bortz.
\newblock Coarse-graining hamiltonian systems using wsindy.
\newblock {\em Scientific Reports}, 14(1):14457, Jun 2024.

\bibitem{KernelLearning5}
J.~Miller, S.~Tang, M.~Zhong, and M.~Maggioni.
\newblock Learning theory for inferring interaction kernels in second-order interacting agent systems.
\newblock {\em Sampling Theory, Signal Processing, and Data Analysis}, 21(1):21, 2023.

\bibitem{Onsager1}
L.~Onsager.
\newblock Reciprocal relations in irreversible processes. {I}.
\newblock {\em Physical Review}, 37:405--426, 1931.

\bibitem{Onsager2}
L.~Onsager.
\newblock Reciprocal relations in irreversible processes. {II}.
\newblock {\em Physical Review}, 38:2265--2279, 1931.

\bibitem{VI4SDE}
M.~Opper.
\newblock Variational inference for stochastic differential equations.
\newblock {\em Annalen der Physik}, 531(3):1800233, 2019.

\bibitem{NFs2}
G.~Papamakarios, E.~Nalisnick, D.~J. Rezende, S.~Mohamed, and B.~Lakshminarayanan.
\newblock Normalizing flows for probabilistic modeling and inference.
\newblock {\em arXiv preprint arXiv:1912.02762v1}, 2019.

\bibitem{KDE2}
E.~Parzen.
\newblock {On Estimation of a Probability Density Function and Mode}.
\newblock {\em The Annals of Mathematical Statistics}, 33(3):1065--1076, 1962.

\bibitem{PINN}
M.~Raissi, P.~Perdikaris, and G.~Karniadakis.
\newblock Physics-informed neural networks: A deep learning framework for solving forward and inverse problems involving nonlinear partial differential equations.
\newblock {\em Journal of Computational Physics}, 378:686--707, 2019.

\bibitem{NFs1}
D.~J. Rezende and S.~Mohamed.
\newblock Variational inference with normalizing flows.
\newblock {\em arXiv preprint arXiv:1505.05770v6}, 2016.

\bibitem{Yannis1}
R.~Rico-Martinez, J.~Anderson, and I.~Kevrekidis.
\newblock Continuous-time nonlinear signal processing: a neural network based approach for gray box identification.
\newblock In {\em Proceedings of IEEE Workshop on Neural Networks for Signal Processing}, pages 596--605, 1994.

\bibitem{risken1996fokker}
H.~Risken and H.~Risken.
\newblock {\em Fokker-planck equation}.
\newblock Springer, 1996.

\bibitem{KDE1}
M.~Rosenblatt.
\newblock {Remarks on Some Nonparametric Estimates of a Density Function}.
\newblock {\em The Annals of Mathematical Statistics}, 27(3):832--837, 1956.

\bibitem{russo1993forward}
F.~Russo and P.~Vallois.
\newblock Forward, backward and symmetric stochastic integration.
\newblock {\em Probability theory and related fields}, 97:403--421, 1993.

\bibitem{PINNinverse1}
H.~Schaeffer.
\newblock Learning partial differential equations via data discovery and sparse optimization.
\newblock {\em Proceedings of the Royal Society A: Mathematical, Physical and Engineering Sciences}, 473(2197):20160446, 2017.

\bibitem{schmidt2009distilling}
M.~Schmidt and H.~Lipson.
\newblock Distilling free-form natural laws from experimental data.
\newblock {\em science}, 324(5923):81--85, 2009.

\bibitem{DGM}
J.~Sirignano and K.~Spiliopoulos.
\newblock Dgm: A deep learning algorithm for solving partial differential equations.
\newblock {\em Journal of computational physics}, 375:1339--1364, 2018.

\bibitem{Rayleigh}
J.~W. Strutt.
\newblock Some general theorems relating to vibrations.
\newblock {\em Proceedings of the London Mathematical Society}, s1-4(1):357--368, 1871.

\bibitem{EnVarA_review}
Y.~Wang and C.~Liu.
\newblock {Some recent advances in energetic variational approaches}.
\newblock {\em Entropy}, 24(5), 2022.

\bibitem{EDMD}
M.~Williams, I.~Kevrekidis, and C.~Rowley.
\newblock A data--driven approximation of the {Koopman} operator: Extending dynamic mode decomposition.
\newblock {\em Journal of Nonlinear Science}, 25(6):1307--1346, 2015.

\bibitem{Kernel4PDEs}
Z.~Xu, D.~Long, Y.~Xu, G.~Yang, S.~Zhe, and H.~Owhadi.
\newblock Toward efficient kernel-based solvers for nonlinear {PDE}s.
\newblock In {\em Forty-second International Conference on Machine Learning}, 2025.

\bibitem{kernelFlow1}
L.~Yang, X.~Sun, B.~Hamzi, H.~Owhadi, and N.~Xie.
\newblock Learning dynamical systems from data: A simple cross-validation perspective, part {V}: Sparse kernel flows for 132 chaotic dynamical systems.
\newblock {\em Physica D: Nonlinear Phenomena}, 460:134070, 2024.

\bibitem{OnsagerNet}
H.~Yu, X.~Tian, W.~E, and Q.~Li.
\newblock Onsagernet: Learning stable and interpretable dynamics using a generalized onsager principle.
\newblock {\em Phys. Rev. Fluids}, 6:114402, 2021.

\bibitem{dissipativeNN2}
J.~Zhang, S.~Zhang, J.~Shen, and G.~Lin.
\newblock Energy-dissipative evolutionary deep operator neural networks.
\newblock {\em Journal of Computational Physics}, 498:112638, 2024.

\bibitem{BiLOforPDE}
R.~Z. Zhang, X.~Xie, and J.~S. Lowengrub.
\newblock Bilo: Bilevel local operator learning for pde inverse problems.
\newblock {\em arXiv preprint arXiv:2404.17789}, 2024.

\bibitem{zhang2022gfinns}
Z.~Zhang, Y.~Shin, and G.~Em~Karniadakis.
\newblock Gfinns: Generic formalism informed neural networks for deterministic and stochastic dynamical systems.
\newblock {\em Philosophical Transactions of the Royal Society A}, 380(2229):20210207, 2022.

\end{thebibliography}

\end{document}